\documentclass[preprint,12pt]{elsarticle}

\usepackage{amssymb}

\usepackage{amsmath,amssymb}
\usepackage{graphicx}
\usepackage{caption}
\usepackage{subcaption}
\usepackage{float}
\usepackage{multirow}

\begin{document}

\begin{frontmatter}


\ead{georgina.al-badri@ucl.ac.uk}

\title{Formation of vascular-like structures using a chemotaxis-driven multiphase model}


\author[inst1,inst2]{Georgina Al-Badri}
\author[inst2,inst3]{James B. Phillips}
\author[inst2,inst4]{Rebecca J. Shipley}
\author[inst1,inst2]{Nicholas C. Ovenden}

\affiliation[inst1]{organization={Department of Mathematics},
            addressline={University College London}, 
            city={London},
            country={UK}}

\affiliation[inst2]{organization={Centre for Nerve Engineering},
            addressline={University College London}, 
            city={London},
            country={UK}}
            
\affiliation[inst4]{organization={Department of Mechanical Engineering},
            addressline={University College London}, 
            city={London},
            country={UK}}

\affiliation[inst3]{organization={Department of Pharmacology},
            addressline={University College London}, 
            city={London},
            country={UK}}

\begin{abstract}
We propose a continuum model for pattern formation, based on the multiphase model framework, to explore \textit{in vitro} cell patterning within an extracellular matrix. We demonstrate that, within this framework, chemotaxis-driven cell migration can lead to formation of cell clusters and vascular-like structures in 1D and 2D respectively. The influence on pattern formation of additional mechanisms commonly included in multiphase tissue models, including cell-matrix traction, contact inhibition, and cell-cell aggregation, are also investigated. Using sensitivity analysis, the relative impact of each model parameter on the simulation outcomes is assessed to identify the key parameters involved. Chemoattractant-matrix binding is further included, motivated by previous experimental studies, and to augment the spatial scale of patterning to within a biologically plausible range. Key findings from the in-depth parameter analysis of the 1D models, both with and without chemoattractant-matrix binding, are demonstrated to translate well to the 2D model, obtaining vascular-like cell patterning for multiple parameter regimes. Overall, we demonstrate a biologically-motivated multiphase model capable of generating long-term pattern formation on a biologically plausible spatial scale both in 1D and 2D, with applications for modelling \textit{in vitro} vascular network formation.
\end{abstract}



\begin{keyword}
multiphase modelling \sep vasculogenesis \sep chemotaxis \sep tissue engineering 
\PACS 0000 \sep 1111
\MSC 0000 \sep 1111
\end{keyword}

\end{frontmatter}


\section{Introduction}
\label{sec:intro}

\textit{In vitro} vascularisation of 3D engineered tissues is required for a range of applications, including \textit{in vitro} models of development and disease e.g.~vascularised tumours \cite{Bittner2020}, and to support the rapid development of blood supply to engineered tissues implanted \textit{in vivo}, e.g.~in a thick skin graft \cite{Tremblay2005, Shen2016}. Development of new strategies to fabricate prevascularised engineered tissues is ongoing, and is commonly described as one of the key challenges facing clinical application \cite{Song2018, Yang2020, Rademakers2019, Chang2017}. As engineered tissues are a complex combination of material properties and cell types, fabricated under a wide range of culture conditions, understanding the cell-level mechanisms behind \textit{in vitro} vascular network formation and the interactions between network-forming endothelial cells and other therapeutic cell types is essential to inform fabrication techniques. 

Mathematical modelling of cell patterning, including of \textit{in vitro} vascular network formation, offers a valuable tool to aid in the identification of the key cues and mechanisms involved \cite{Scianna2013}, and to enable \textit{in silico} trial and error of different experimental variables \cite{Waters2021}. Where individual cell cues and mechanisms are difficult to isolate and investigate experimentally, mathematical modelling offers the opportunity to probe the influence of each mechanism individually. Furthermore, computational techniques such as sensitivity analysis offer a powerful and cost-effective way to assess the relative importance of cell mechanisms and experimental variables.

A first step to using simulation to guide experimental methods is to select a suitable mathematical framework that is capable of mimicking the behaviours observed \textit{in vitro}. Here, a continuous multiphase model framework is proposed that has not been previously utilised to explore cell pattern formation within a 3D environment. Crucially, the multiphase model framework can resolve both global boundary conditions and cell-level chemical and mechanical cues, in order to fully support the mechanisms associated with the migration of a cell phase within a 3D construct (such as a hydrogel or scaffold). The multiphase model framework has been utilised in several studies of \textit{in vitro} tissue culture and growth modelling \cite{ODea2012}, and was first adapted for this application by Lemon et al.~\cite{Lemon2006, Lemon2007}. A multiphase model approach has also been used to study tumour growth \cite{Byrne2003, Preziosi2009, Tosin2010, Hubbard2013, Sciume2013}. More recently, Dyson et al.~\cite{Dyson2016} extended the multiphase model framework to consider in more detail the mechanical effects of a fibrous collagen scaffold on cell migration and tissue growth. The multiphase model framework has additionally been demonstrated to capture chemotaxis-driven pattern formation, with cell aggregation demonstrated in 1D, and in a 2D thin film limit \cite{Green2017}. 

The behaviours and responses of endothelial cells (ECs), which can form vascular-like networks \textit{in vitro}, have been investigated in a wide-range of \textit{in vitro} assays \cite{Bayless2000, Koh2008, Stratman2009, Blinder2016}. Vascular endothelial growth factor (VEGF) is known to be essential for vascular network formation from knockout \textit{in vivo} mouse studies that showed deletion of the gene responsible for VEGF production led to improper blood vessel formation and lethality \cite{Carmeliet1996, Ferrara1996}. The mechanisms by which VEGF may facilitate endothelial network formation are varied, including (i) accelerating the rate of cell migration by chemokinesis \cite{Zetter1980}, (ii) influencing the direction of cell migration as a chemoattractant \cite{Barkefors2008, Wu2014}, (iii) affecting the rate of endothelial cell proliferation \cite{Unemori1992, Seghezzi1998}, and (iv) inducing/increasing cell production of matrix metalloproteinases (MMPs), enzymes responsible for degrading ECM proteins in the matrix surrounding the cell \cite{Unemori1992, Hanjaya-Putra2010}. 

\textit{In vitro} studies have found that HUVECs (human umbilical vein ECs) can form vascular networks alone without the addition of exogenous VEGF, particularly under low oxygen conditions  \cite{Helmlinger2000, Serini2003}. Serini et al.~\cite{Serini2003} measured detectable levels of VEGF produced by both HUVECs (0.24 ng/ml) and microvascular ECs (0.67 ng/ml) after 3 hours. Moreover, the role of VEGF gradients for EC network formation was confirmed by observing inhibition of network formation via the addition of either anti-VEGF, or via a saturating amount of exogenous VEGF \cite{Serini2003}. 


To accurately model the dynamics of VEGF, the bioactivity of the scaffold material must also be considered. There is substantial evidence that some extracellular matrix proteins can bind soluble proteins, e.g. growth factors \cite{Macri2007}. In particular, it has been shown that fibronectin binds with VEGF \cite{Wijelath2006}, as does fibrin \cite{Sahni2000}, and heparin \cite{Ruhrberg2002}. Kohn-Luque et al.~\cite{Kohn-Luque2013} demonstrated that exogenously added VEGF could bind to Matrigel pericellularly, and that this binding was co-localised with specific proteins including cell-secreted fibronectin and heparin. The authors further demonstrated that matrix-bound VEGF elicited a stronger chemotactic response in endothelial cells than soluble VEGF, and matrix-bound VEGF was protected from degradation \cite{Kohn-Luque2013}.


Previous partial differential equation (PDE)-based vasculogenesis models have been predominantly based on a 2D assay in which endothelial cells are seeded on, as opposed to within, the extracelllar matrix \cite{Scianna2013}. Notable models were able to replicate some of the resulting network properties, with similar qualitative behaviours as the initial cell density was varied \cite{Scianna2013, Serini2003, Tosin2006}. In addition, the spatial scale of the \textit{in vitro} network, determined by intervessel separation, was observed to be of the order of 200 $\mu$m and well matched by the numerical simulations \cite{Serini2003}. However, key mechanisms, such as cell-matrix traction were not included \cite{Serini2003}, and often vascular-like patterning was observed transiently \cite{Tosin2006, Manoussaki1996, Tranqui2000, Namy2004}. To date, there is an absence of a continuous model of vasculogenesis with a focus on mechanisms relevant to 3D culture, in which ECs are seeded within a scaffold or hydrogel. Crucially, existing continuous models have not yet been able to generate stable vascular-like structures.

In this paper, the impact of several commonly proposed mechanisms for vascular network formation are explored within the multiphase model framework; these are assessed using computational techniques including sensitivity analyses and parameter optimisation. Long-term cell cluster formation is demonstrated, and a model form including VEGF-matrix binding is shown to form cell clusters on a comparable spatial scale to \textit{in vitro} vascular network pattern size and structure. 

In Section \ref{sec:model}, a general multiphase model is presented, prior to the selection of specific constitutive functional forms for intraphase cell forces and solute reaction terms. The resulting 2D model is then nondimensionalised in Section \ref{sec:dim_analysis} using parameter values motivated by literature. A brief summary of the procedure for the computational implementation of the model in Python is then given in Section \ref{sec:comp_methods}, followed by an outline of the computational analyses methods selected. Results relating to the impact of initial conditions, core model parameters, and additional common mechanisms are presented in Sections \ref{res:init_cond_analysis} - \ref{res:add_mech}. In Section \ref{res:vegf_binding}, the impact of the addition of VEGF-matrix binding to the model is considered and shown to aid pattern formation on a spatial scale relevant to vascular network formation \textit{in vitro}. The difference in spatial scale generated by the core and binding models in 1D and 2D is further demonstrated in Section \ref{res:spac_anal_all}.


\section{Model outline}
\label{sec:model}
The general governing equations for the multiphase model are derived following constitutive forms proposed by Lemon et al.~\cite{Lemon2006} and subsequent model development papers \cite{Lemon2007, ODea2009, Pearson2014}. Following a discussion of possible functional forms for reaction terms, and parameter value ranges motivated by literature, the resulting coupled PDE model system is nondimensionalised for a 2D Cartesian geometry. 

\subsection{General equations}
\label{model:gen_eqns}
Three phases are modelled to form a cellular hydrogel: endothelial cells, water (culture media) and extracellular matrix (the hydrogel's solid component), for which volume fractions are denoted by $n$, $w$, and $m$ respectively. In the absence of matrix degradation or deposition, the extracellular matrix volume fraction $m$ is assumed to be uniform and constant. A no-voids constraint, $n + w + m = 1$, is imposed,
and a constant $\phi = 1 - m$ is introduced, such that $n + w = \phi$.

The general conservation of mass equations governing the cell volume fraction ($n$), and water volume fraction ($w$), take the form of an advection-diffusion equation and advection equation respectively

\begin{equation}\label{eqnnmass}
\frac{\partial n}{\partial t} + \nabla \cdot \left( n \mathbf{u_n} \right) = \nabla \left(D_n \nabla n \right),
\end{equation}
and
\begin{equation}\label{eqnwmass}
\frac{\partial w}{\partial t} + \nabla \cdot \left( w \mathbf{u_w} \right) = 0,
\end{equation}
where $D_n$ is the cell diffusion rate, and $\mathbf{u_i}$ is the velocity of each phase. We focus on the most commonly used hydrogel material, collagen, and hence assume that cell proliferation is negligible as observed during vascular network formation within a collagen hydrogel \cite{Helmlinger2000, Odedra2011}. The general conservation of mass equation for a solute with concentration $c$ takes the form of the following advection-diffusion-reaction equation,
\begin{equation}\label{eqncgen}
\frac{\partial (cw)}{\partial t} + \nabla \cdot \left( c w \mathbf{u_w} \right) = \nabla \cdot \left( D_c w \nabla c \right) + R,
\end{equation}
where $D_c$ is the diffusion coefficient, and $R$ is a reaction term that can be a function of $n$, $w$, and/or $c$. \\

Conservation of momentum equations include all the intraphase (internal) and interphase (external) forces. Due to the relatively short spatial scale (typically millimetres), in comparison to a long timescale for cell motility (typically hours - see Table \ref{tab:timescales}), inertial effects are relatively very small and hence can be neglected. The general conservation of momentum equations for the cell phase ($n$) and water phase ($w$) are therefore represented by a balance between intraphase stress and external forces resulting from the pressure between phases,
\begin{equation}\label{eqnmomn}
\nabla \cdot \left( n \mathbf{\sigma}_n \right) + \mathbf{f}_{nw} +  \mathbf{f}_{nm} = 0,
\end{equation}
and
\begin{equation}\label{eqnmomw}
\nabla \cdot \left( w \mathbf{\sigma}_w \right) +  \mathbf{f}_{wn} +  \mathbf{f}_{wm} = 0.
\end{equation}
Here, $\sigma_i$ is the stress tensor for phase $i$, and $\mathbf{f}_{ij}$ represents the interphase force of phase $j$ on phase $i$ noting that, by Newton's third law, $\mathbf{f}_{ij} = - \mathbf{f}_{ji}$. Constitutive forms of the stress tensors and interphase forces are prescribed in the following sections following Lemon et al.~\cite{Lemon2006}.

\subsubsection{Interphase forces}
\label{model:interphase_forces}
The interphase forces are split into two components: interphase pressure, $p_{ij}$, and drag, $\gamma_{ij}$. The force exerted by phase $j$ on  phase $i$ is thus given by 
\begin{align}\label{interphase}
\mathbf{f}_{ij} &= p_{ij} j \nabla i - p_{ji} i \nabla j + \gamma_{ij} i j \left(\mathbf{u}_j - \mathbf{u}_i \right) 
.
\end{align}
Here, the interphase pressures $p_{ij}$ are decomposed into a `contact-independent' pressure equal to the water pressure, $p_w$, and additional `active' pressures. We consider only a traction-induced pressure force between the cells and extracellular matrix $\psi_{nm}$, i.e $p_{nm} = p_{w} + \psi_{nm}$. This traction force is further assumed to be a (zero or negative) constant, such that $\psi_{nm} = - \eta$. Based on these assumptions, each interphase force term can now be expressed as
\begin{eqnarray*}
\mathbf{f}_{nw} &=& p_w w \nabla n - p_w n \nabla w + \gamma_{nw} nw\left(\mathbf{u_w} - \mathbf{u_n} \right), \\
\mathbf{f}_{nm} &=& (p_{w} - \eta) m \nabla n - \gamma_{nm} nm \mathbf{u_n}, \\
\mathbf{f}_{wm} &=& p_w m \nabla w - \gamma_{wm} w m \mathbf{u_w}. 
\end{eqnarray*}

\subsubsection{Stress tensors}
\label{model:stress_tensors}
The cell phase is modelled as a viscous liquid, with stress tensor $$\sigma_n = -p_n \mathbf{I} + \mu_n \mathbf{\tau}_n.$$ The cell intraphase pressure, $p_n$, is similarly split into a water pressure component, $p_w$, an additional term to account for cell-cell interactions, and a final term to account for cell-matrix interactions, such that
$$p_n = p_w + \Pi + m \psi_{nm}. $$ 
The cell-cell interactions, denoted $\Pi$, can include: (i) contact inhibition, e.g. $\Pi_n = \delta_n n^2/(\phi - n)$ \cite{ODea2009}; (ii) cell aggregation, e.g. $\Pi_a = - \nu n$; and (iii) chemotaxis, e.g. $\Pi_c = \chi \exp \left( - \frac{c}{c_M} \right) $~\cite{Pearson2014}.
Chemotaxis is considered to be the dominant cell intraphase pressure and included from the outset; the possible influence of contact inhibition and cell-cell aggregation is explored subsequently in Section \ref{res:add_mech}. 

The water phase is modelled as an inviscid fluid such that $\sigma_w = -p_w \mathbf{I}$, where $p_w$ is the water pressure; we therefore assume that there are no other contributing terms due to intraphase or interphase pressures. 

Substituting these terms into (\ref{eqnmomn}) and (\ref{eqnmomw}), and applying the no-voids constraint, the conservation of momentum equation for the cell phase becomes
\begin{equation}\label{simpmomn}
- n \nabla p_w - \nabla \left( n \Pi \right) + \nabla \left(n \mu_n \tau_n \right) +  \gamma_{nw} nw\left(\mathbf{u_w} - \mathbf{u_n} \right) - \gamma_{nm} nm \mathbf{u_n} - \eta m \nabla n = 0,
\end{equation}
and for the water phase we have
\begin{equation}\label{simpmomw}
-w \nabla p_w - \gamma_{nw} nw\left(\mathbf{u_w} - \mathbf{u_n} \right) - \gamma_{wm} wm\mathbf{u_w}  = 0.
\end{equation}

\subsubsection{Solute dynamics}
\label{model:vegf_reaction}
The main VEGF dynamics are driven by its production and uptake by the endothelial cells, along with cell-independent degradation. The reaction term for VEGF concentration using (\ref{eqncgen}) is thus written 
$$ R(n, w, c) = \alpha(n, w, c) - \kappa(n, w, c) - \delta c, $$
where $\delta$ is the constant degradation rate, $\alpha(n, w, c)$ is a production function, and $\kappa(n, w, c)$ is an uptake function, with possible dependencies indicated. Although uptake of VEGF, which spans the ligand-receptor binding and internalisation process of VEGF into the cell, is required for a cell response, in this case we decouple the process and take uptake to represent the removal of VEGF from the surrounding fluid only.   

A Michaelis-Menten function of the form \cite{Holmes2000}
$ \kappa(n, w, c) = \kappa n w c/(c + K) $
has previously been used to describe cellular uptake of VEGF in an angiogenesis model,
where $\kappa$ is the maximal uptake rate, and $K$ is the concentration of VEGF at which uptake is half-maximal. 
A Michaelis-Menten term has also been used to model production, for example by Holmes and Sleeman \cite{Holmes2000} for fibronectin production by endothelial cells during angiogenesis. As the local concentration of fibronectin increases, the rate of production slows, until it reaches a maximum rate. Alternatively, in a fluids-based mechano-chemical model of vasculogenesis, Tosin et al.~\cite{Tosin2006} consider a production rate $\alpha(n)$ with non-linear dependence on the cell density. The authors assumed a production rate that reached a finite maximum for $n=0$ and is monotonically decreasing for $n>0$, reaching a negligible level once the local cell density represents a sizable cluster. Following Tosin et al.~\cite{Tosin2006}, we select the function form $ \alpha(n) = \alpha n (\phi - n)$ that exhibits the same behaviour when considering cell volume fraction as opposed to cell density.

Including the reaction terms, the governing equation for the solute VEGF is therefore
\begin{equation}\label{eqnc}
\frac{\partial (cw)}{\partial t} + \nabla \cdot \left( c w \mathbf{u_w} \right) = \nabla \cdot \left( D_c w \nabla c \right) + \alpha n (\phi - n) - \frac{\kappa n w c}{c+ K} - \delta c.
\end{equation}
The governing equations are now (\ref{eqnnmass}, \ref{eqnwmass}, \ref{simpmomn} - \ref{eqnc}). These are combined with suitable boundary conditions dependent on the geometry, and initial conditions dependent on the experimental setup as described in the next section. The variable and parameter definitions for the general and specified models are given in Table~\ref{tab:generalparams}.

\begin{table}[H]
\setlength{\tabcolsep}{10pt} 
\renewcommand{\arraystretch}{1.5} 
\begin{tabular}{ l l }
 Variable/parameter & Definition \\ 
 \hline
 \hline
 n & cell volume fraction \\  
 w & water volume fraction \\
 m & matrix volume fraction \\
 $\phi = 1 - m$ & available volume (constant) \\
 c & chemical solute concentration \\
 $\mathbf{u_i}$ = ($u_i$, $v_i$) ($i$ = n, w)  & cell/water velocity \\
 $D_i$ ($i$ = n, c) & diffusion rate of cells/chemical solute \\
 $R$ & reaction term of chemical solute \\
 $\sigma_i$ (i = n, w) & cell/water stress tensor \\
 $f_{ij}$ ($i$, $j$ = n, w, m, $i \neq j$) & interphase force exerted on phase $i$ by phase $j$ \\
 $p_i$ (i = n, w) & pressure in cell/water phase \\
 $\mu_i$ ($i$ = n, w) & cell/water viscosity \\
 $\tau_i$ ($i$ = n, w) & cell/water deviatoric stress tensor \\
 $\delta_i$ & contact inhibition-driven pressure \\
 $\nu_n$ & aggregation pressure \\
 $\chi$ & chemotaxis-driven pressure \\
 $p_{ij}$ ($i$, $j$ = n, w, m, $i \neq j$) & interphase pressure exerted on phase $i$ by phase $j$ \\
 $\gamma_{ij}$ ($i$, $j$ = n, w, m, $i \neq j$) & drag coefficient between phase $i$ and phase $j$ \\
 $\psi_{nm}$ & traction coefficient between cell and matrix phase \\
 $\eta$ & (constant) traction pressure between cell and matrix phase \\
 $\alpha$ & VEGF production rate \\
 $\kappa$ & VEGF uptake rate \\
 $\delta$ & VEGF degradation rate \\
 \hline
\end{tabular}
\caption{Variable and parameter definitions for the general and specified multiphase model.}
\label{tab:generalparams}
\end{table}


\section{Dimensional Analysis}
\label{sec:dim_analysis}
\subsection{Dimensional parameter values}
\label{model:param_values}

\begin{figure}[h]
    \centering
    \includegraphics[width=0.45\textwidth]{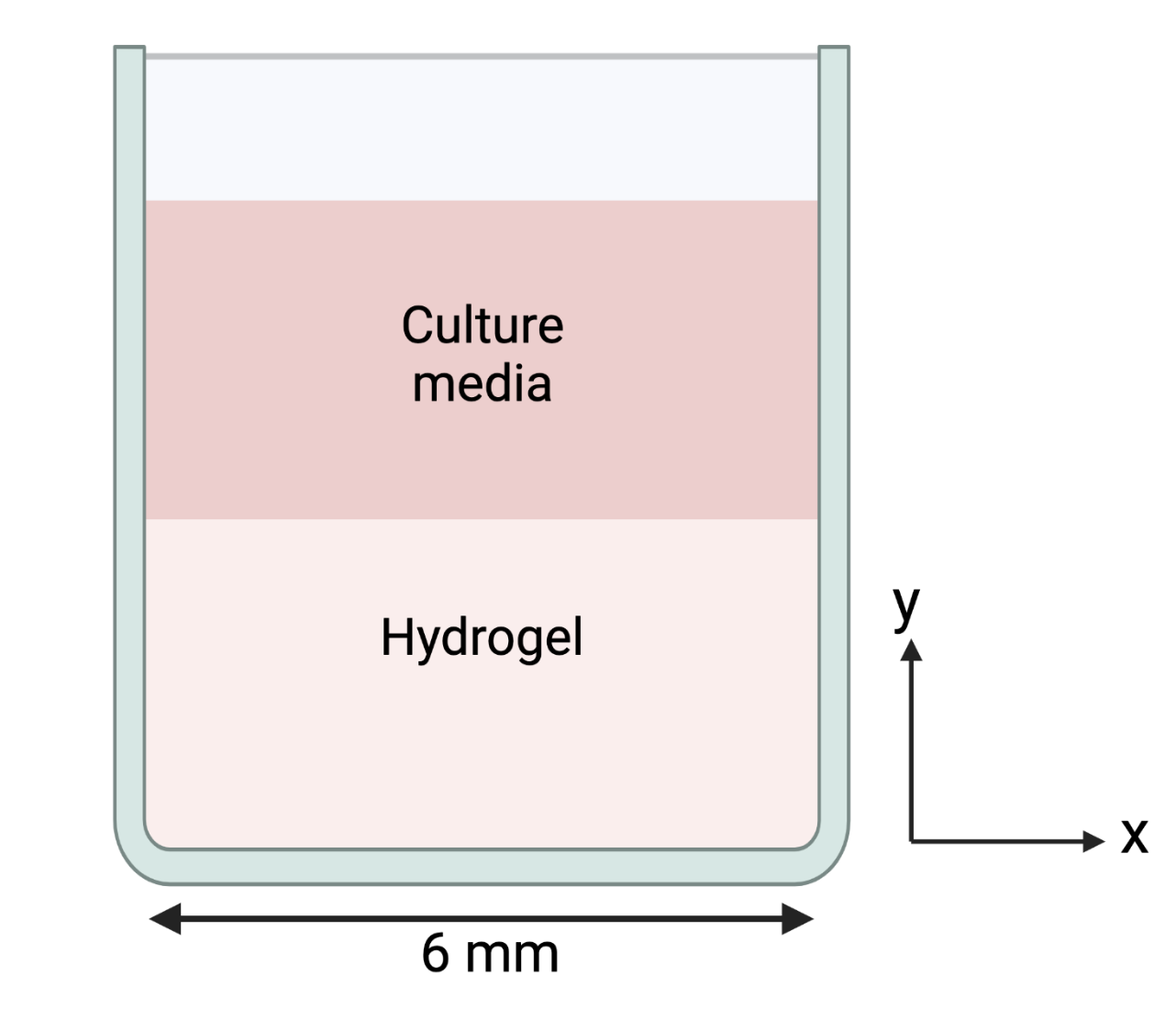}
    \caption{Cylindrical well geometry for a 96-well plate, commonly used for cellular hydrogel assays.}
    \label{fig:2Dgeom}
\end{figure}

The typical length scale is chosen to be based on the dimensions of a 96-well plate, shown in Figure \ref{fig:2Dgeom}, commonly used experimentally for the analysis of cell-seeded hydrogels. Given a diameter of around 6mm, and variable hydrogel height of 1 - 5mm, we set the representative length scale $L$ to be within this range at 3mm. The representative timescale is chosen to be related to the time under which network formation is observed \textit{in vitro}, shown to span from a few hours \cite{Serini2003}, to 24 - 48 h \cite{Holmes2000}, to several days \cite{Paponthesis}, dependent on the geometry and experimental conditions. A timescale of $T = 10^5$ s , $\sim 27$ h, is chosen to match the order of the timescale of chemotaxis-driven pattern formation, which is the focus of the model presented here.

Table~\ref{tab:timescales} outlines the estimated timescales for the main processes considered by the model. Excluding cell diffusion, each process considered is clearly significant on this length scale given a timescale of interest of 24 - 96 hours. Cell diffusion is maintained to improve numerical model stability. The remaining cell- and solute-related parameters required for the model are discussed below. 

\begin{table}[H]
\setlength{\tabcolsep}{10pt} 
\renewcommand{\arraystretch}{1.5} 
\begin{tabular}{ l l l}
\hline
Process & Estimated timescale & Reference(s) \\
\hline

Cell diffusion & 2.5 $\times$ 10$^3$ - 2.5 $\times$ 10$^4$ h & \cite{Stokes1991, Rupnick1988, Anderson1998c}\\

VEGF diffusion in a hydrogel & 250 h & \cite{Serini2003, Kohn-Luque2013, Nunez2006} \\

VEGF diffusion in media & 25 h & \cite{MacGabhann2007} \\

VEGF production & 0.05 h & \cite{Serini2003} \\

VEGF half-life & 0.1 h - 3 h & \cite{Kohn-Luque2011, Serini2003, Merks2006, Chen2010, Kohn-Luque2013} \\

Vascular network formation & 9 h - 48 h+ & \cite{Serini2003, Holmes2000, Helmlinger2000}\\

Cell directed migration in a hydrogel & 80 h & \\

\hline
\end{tabular}
\caption{Estimated timescales of relevant dynamic processes. Estimates are based on the hydrogel length scale L = 3 mm, and a typical concentration of VEGF, 1 ng/ml. References are given for each parameter where applicable.}
\label{tab:timescales}
\end{table}

\subsubsection{Cell-related parameters}
\label{model:cell_params}
Using Stokes flow, Lemon and King \cite{Lemon2007} estimated the magnitude of the water-matrix drag and cell-water drag to be comparable, and the cell-matrix drag term  to be several orders of magnitude larger. Hence, we take the cell-water drag $\gamma_{nw}$ term in Equation \ref{eqn:2Dun} to be negligible in comparison to the other mechanisms. However, to maintain a relationship between the water pressure and water velocity in Equation (\ref{simpmomw}), we must maintain the water-matrix drag term $\gamma_{wm}$. The estimate of cell viscosity, $\mu_n$ = 1 $\times$ 10$^4$N m$^{-2}$ s, is based on experimental evidence also discussed by Lemon and King \cite{Lemon2007}

The strength of chemotaxis, denoted $\chi$, is modelled as a pressure force within this framework and is hence not amenable to experimental derivation. We explore possible dimensionless values of $\chi$ such that chemotactic effects are dominant. The cell-matrix traction constant $\eta$ is explored in a similar fashion.

\subsubsection{VEGF-related parameters}
\label{model:vegf_params}
There is scant quantitative data regarding the concentration of VEGF produced by endothelial cells: Serini et al.~\cite{Serini2003} measured detectable levels of VEGF after 3 hours in the range 0.24 ng/ml - 0.67 ng/ml; similarly Nomura et al.~\cite{Nomura1995} quoted that HUVECs produced a concentration of 0.15 ng/ml at 10\% oxygen. Based on these estimates, the concentration of 1 ng/ml is taken as a typical scaling for VEGF concentration, $c_M$.

Parameters including VEGF production and uptake rates are highly dependent on cell type, and again scant quantitative data can be found. VEGF production rates used in previous models in literature are often quoted as per area per second, reflecting the 2D \textit{in vitro} setup of the assay being modelled. A crude conversion made from the estimate by K\"ohn-Luque et al.~\cite{Kohn-Luque2011}, combined with comparison of production rates from other cell types in 3D environments such as ADSCs \cite{Coy2020}, can be used to estimate a dimensional range of  $\alpha$ from $2 \times 10
^{-13}$ to $2 \times 10^{-11}$ g ml$^{-1}$ s$^{-1}$. However, parameter analysis methods are required here to determine a suitable range of values that may lead to qualitatively different model outcomes, including the parameter sweep presented in Section \ref{res:core_analysis}.


\subsection{Nondimensionalisation}
\label{model:nondimensionalisation}

Based on Figure \ref{fig:2Dgeom}, a 2D Cartesian model is considered to approximate a top-down view of the cylindrical well geometry. The geometry is considered to be square for simplicity. The following dimensionless variables are used for nondimensionalisation of the 2D Cartesian model, denoted by an asterisk (*):
\begin{align*}
x^* &= \frac{x}{L}, & y^* &= \frac{y}{L}, & t^*  &= \frac{t}{T}, & c^* &= \frac{c}{c_M}, \\
D_i^* &= \frac{D_i T}{L^2}, & u_{n, w}^* &= \frac{u_{n, w} T}{L}, & v_{n, w}^* &= \frac{v_{n, w} T}{L}, & \gamma^*_{ij} &= \frac{\gamma_{ij} L^2}{\mu_n}, \\
\chi^* &= \frac{\chi T}{\mu_n}, & \eta^* &= \frac{\eta T}{\mu_n}, & p_w^* &= (p_w - p_{atm})\frac{T}{\mu_n}, & \alpha^* & = \frac{\alpha T}{c_M}, \\
\kappa^* &= \frac{\kappa T}{c_M}, & K^* & = \frac{K}{c_M}, & \delta^* &= \delta T.
\end{align*}

After nondimensionalisation, dropping asterisks, the 2D Cartesian model comprises eight coupled PDEs. The following dimensionless equations govern cell volume fraction ($n$), water volume fraction ($w$), water velocity components ($u_w$, $v_w$), cell velocity components ($u_n$, $v_n$), water pressure ($p_w$), and VEGF concentration ($c$): 
\begin{equation}
\label{eqn:2Dn}
\frac{\partial n}{\partial t} +\frac{\partial}{\partial x}\left(n u_{n}\right) +\frac{\partial}{\partial y}\left(n v_{n}\right) =  D_n \left( \frac{\partial^{2} n}{\partial x^{2}} + \frac{\partial^{2} n}{\partial y^{2}} \right),
\end{equation}

\begin{equation}
\label{eqn:novoids}
    n + w = \phi,
\end{equation}

\begin{equation}
\label{eqn:2Dpw}
    \frac{\partial w}{\partial t} -\frac{1}{\gamma_{w m} m} \frac{\partial}{\partial x}\left(w \frac{\partial p_{w}}{\partial x}\right) -\frac{1}{\gamma_{w m} m} \frac{\partial}{\partial y}\left(w \frac{\partial p_{w}}{\partial y}\right)=0,
\end{equation}

\begin{equation}
\label{eqn:2Duw}
    u_{w} = - \frac{1}{\gamma_{w m} m} \frac{\partial p_{w}}{\partial x},
\end{equation}

\begin{equation}
\label{eqn:2Dvw}
    v_{w} = - \frac{1}{\gamma_{w m} m} \frac{\partial p_{w}}{\partial y},
\end{equation}

\begin{multline}
\label{eqn:2Dc}
 \frac{\partial(c w)}{\partial t} + \frac{\partial(c w u_w)}{\partial x} + \frac{\partial(c w v_w)}{\partial y} = \frac{\partial}{\partial x} \left(D_c w \frac{\partial c}{\partial x} \right) + \frac{\partial}{\partial y} \left(D_c w \frac{\partial c}{\partial y} \right) \\
 + \alpha n (\phi - n) - \frac{\kappa n\, w\, c}{K + c} - \delta c,
\end{multline}

\begin{multline}\label{eqn:2Dun}
- n\frac{\partial p_w}{\partial x}  - \frac{\partial}{\partial x}\left( n \chi \exp(-c) - \nu n^2 + \frac{\delta_n n^3}{\phi - n} \right) + \frac{2}{3} \frac{\partial}{\partial x} \left[ n \left( 2\frac{\partial u_n}{\partial x} - \frac{\partial v_n}{\partial y} \right) \right]  \\
+ \frac{\partial}{\partial y}\left[ n \left(\frac{\partial u_n}{\partial y} + \frac{\partial v_n}{\partial x} \right) \right] 
- \gamma_{nm} n m u_n - \eta m \frac{\partial n}{\partial x} = 0,
\end{multline}

\begin{multline}\label{eqn:2Dvn}
- n\frac{\partial p_w}{\partial y}  - \frac{\partial}{\partial y}\left( n \chi \exp(-c) - \nu n^2 + \frac{\delta_n n^3}{\phi - n} \right) + \frac{\partial}{\partial x} \left[ n \left( \frac{\partial u_n}{\partial y } + \frac{\partial v_n}{\partial x} \right) \right] \\
+ \frac{2}{3}\frac{\partial}{\partial y}\left[ n \left(2\frac{\partial v_n}{\partial y} - \frac{\partial u_n}{\partial x} \right) \right] 
- \gamma_{nm} n m v_n - \eta m \frac{\partial n}{\partial y}  = 0.
\end{multline}


%











\subsection{Boundary conditions}
\label{model:boundary_conds}
No flux conditions are applied at all boundaries on the cell phase, water phase (including water pressure), and the VEGF concentration. No slip conditions are also applied to the cell phase at edges of the domain. Thus we have
\[
 u_{n}=0, \quad v_{n}=0, \quad \frac{\partial n}{\partial x}=0, \quad \frac{\partial c}{\partial x}=0, \quad \frac{\partial p_w}{\partial x}=0, \quad \textrm{at $x=0$, $1$, and} \]
 
 \[u_{n}=0, \quad v_{n}=0, \quad \frac{\partial n}{\partial y}=0, \quad \frac{\partial c}{\partial y}=0, \quad \frac{\partial p_w}{\partial y}=0, \quad \textrm{at $y=0$, $1$.}
\]

Equivalent no flux and no slip boundary conditions are used when considering a 1D version of the model, used in order to exploit computationally expensive analysis methods. 

\subsection{Initial conditions}
\label{model:init_conds}
An initial cell distribution at rest is assumed within the hydrogel, in which no VEGF is present, given that the culture media used for the hydrogel preparation does not contain VEGF. Thus we have
\begin{equation}
\begin{aligned}
  n(x, y, 0) &= n^0(x, y),  \quad  u_n(x, y, 0) = 0, \quad  v_n(x, y, 0) = 0, \\
  p_w(x, y, 0) &= 0, \quad c(x, y, 0) = 0, \quad \textrm{and} \quad  w(x, y, 0) = \phi - n^0(x, y). 
\end{aligned}
\end{equation}

\subsection{Model extension for VEGF-matrix binding}
The addition of a VEGF-matrix binding mechanism introduces three new parameters: the binding rate, $k_b$, the unbinding rate, $k_u$, and the strength of chemotaxis in response to bound VEGF, $\chi_b$. The chemotaxis parameter used in the core model, $\chi$, is renamed $\chi_u$, to indicate that it is the strength of chemotaxis in response to free (unbound) VEGF. In a similar manner, the concentration of unbound VEGF is denoted by $c_u$, and the concentration of bound VEGF by $c_b$. The cell momentum conservation equations balancing intraphase and interphase forces, equations (\ref{eqn:2Dun}, \ref{eqn:2Dvn}), consequently require adaptation to include the chemotactic response to both bound and free VEGF. The conservation of momentum equations for the cell phase in this case take the form
\begin{multline}\label{eqn:2Dun_binding}
- n\frac{\partial p_w}{\partial x}  - \frac{\partial}{\partial x}\left( n \chi_u \exp(-c_u) + n \chi_b \exp(c_b) \right) - \frac{\partial}{\partial x} \left(\nu n^2 + \frac{\delta_n n^3}{\phi - n} \right) \\ 
+ \frac{2}{3} \frac{\partial}{\partial x} \left[ n \left( 2\frac{\partial u_n}{\partial x} - \frac{\partial v_n}{\partial y} \right) \right]  
+ \frac{\partial}{\partial y}\left[ n \left(\frac{\partial u_n}{\partial y} + \frac{\partial v_n}{\partial x} \right) \right] \\
- \gamma_{nm} n m u_n - \eta m \frac{\partial n}{\partial x} = 0,
\end{multline}

\begin{multline}\label{eqn:2Dvn_binding}
- n\frac{\partial p_w}{\partial y}  - \frac{\partial}{\partial y}\left( n \chi_u \exp(-c_u) + n \chi_b \exp(c_b) \right) - \frac{\partial}{\partial y}\left( \nu n^2 + \frac{\delta_n n^3}{\phi - n} \right) \\
+ \frac{\partial}{\partial x} \left[ n \left( \frac{\partial u_n}{\partial y } + \frac{\partial v_n}{\partial x} \right) \right] 
+ \frac{2}{3}\frac{\partial}{\partial y}\left[ n \left(2\frac{\partial v_n}{\partial y} - \frac{\partial u_n}{\partial x} \right) \right] \\
- \gamma_{nm} n m v_n - \eta m \frac{\partial n}{\partial y}  = 0.
\end{multline}

Two advection-reaction-diffusion equations governing the dynamics of $c_u$ and $c_b$ are also required in place of Equation~(\ref{eqn:2Dc}). To consider the simplest model of VEGF binding in the first instance, any diffusion, uptake, or direct production of bound VEGF $c_b$ is neglected. There is some evidence that binding to the extracellular matrix reduces the rate of degradation of proteins including VEGF \cite{Chen2010}; this difference is neglected for now so that the rate of degradation of bound and unbound VEGF is considered equal and represented by parameter $\delta$. The experimental analysis of K\"ohn-Luque et al.~\cite{Kohn-Luque2013} is suggestive of VEGF-binding to specific local cell-deposited ECM proteins such as fibronection. A full model may consider secretion of fibronection, and VEGF binding to this fibronectin only, however to simplify this mechanism here, the pericellular location of the VEGF binding sites can be accounted for by choosing the binding rate to be proportional to the local cell density. 


The equation for unbound VEGF $c_u$, Equation~(\ref{eqn:2Dc}), is adapted to include the transfer of VEGF between the bound and unbound state, and becomes
\begin{multline}
\label{eqn:2Dc_binding}
 \frac{\partial(c_u w)}{\partial t} + \frac{\partial(c_u w u_w)}{\partial x} + \frac{\partial(c_u w v_w)}{\partial y} = \frac{\partial}{\partial x} \left(D_c w \frac{\partial c_u}{\partial x} \right) + \frac{\partial}{\partial y} \left(D_c w \frac{\partial c_u}{\partial y} \right) \\
 + \alpha n (\phi - n) - \frac{\kappa n w c_u}{K + c_u} - \delta c_u - k_b m n c_u w  + k_u w c_b m,
\end{multline}
where the additional terms $k_b n c_u w$ and $k_u w c_b m$ model the rate of binding and unbinding, proportional to the cell volume fraction $n$, and water volume fraction $w$ respectively. The governing equation for bound VEGF concentration, $c_b$, is taken to be
\begin{equation}
\label{eqn:2Dcb}
 \frac{\partial(c_b m)}{\partial t} = k_b m n c_u w  - k_u w c_b m.
\end{equation}

K\"ohn-Luque et al.~\cite{Kohn-Luque2013} fitted experimental data to a similar ODE model of binding and unbinding of VEGF, given by
\begin{equation}
\label{eqn:KLb}
 \frac{\partial(b)}{\partial t} = k_{on}^* U_{eq} - k_{off} b,
\end{equation}
where $b$ is the concentration of bound VEGF, $U_{eq}$ is the equilibrium concentration of unbound VEGF, $k_{on}^*$ is the rate of binding, taking into account the equilibrium concentration of binding sites, and $k_{off}$ is the rate of unbinding. Comparing this with Equation~\ref{eqn:2Dcb}, and using the values of $k_{on}^*$ and $k_{off}$ determined by K\"ohn-Luque et al.~\cite{Kohn-Luque2013}, the values derived for $k_b$ and $k_u$ are $k_b \approx 0.05 \pm 0.01\, \mathrm{s}^{-1}$ and $k_u \approx 3.7 \times 10^{-3} \pm 3.97 \times 10^{-4} \, \mathrm{s}^{-1}$. The impact of the experimental uncertainty will be considered via a sensitivity analysis in Section \ref{res:vegf_binding}.


\section{Computational methods}
\label{sec:comp_methods}

\subsection{Numerical methods}
\label{meth:num_methods}
The 1D and 2D Cartesian models were implemented computationally in Python by first discretising the model equations using the finite difference method. Central-space and backward-time differences were applied, and boundary conditions were applied by adopting a forward- or backward-space difference on the boundary as appropriate. The resulting set of coupled matrix problems were solved algorithmically for each model timestep with use of Python's inbuilt linear algebra solvers. Grid-step and time-step convergence were verified in both cases, and the solvers for each of the 1D model equations were also verified by comparing to exact or approximate analytical solutions where possible.

\subsection{Computational analysis methods}
\label{meth:comp_anal}
To assess the capabilities of the core model to simulate cell patterning, the computational analyses focused on (i) identifying qualitatively distinct cell distribution outcomes, and (ii) assessing the timing and stability of pattern formation.
The overall aim of the analyses was to identify a suitable model that was capable of qualitatively matching \textit{in vitro} data, including the temporal and spatial scale of pattern formation. Additionally, the analyses sought to minimise the number of parameters required in the model to meet these conditions. To exploit computational techniques whilst minimising computational expense, initial analyses were conducted on the 1D version of Equations (\ref{eqn:2Dn} -- \ref{eqn:2Dvn}) and binding extension, assuming no $y$-dependence. Such model reduction is justified as the 1D model retains the likely mechanisms of cluster formation. Indeed, the conclusions of these 1D analyses are shown to translate well to the 2D model throughout Section \ref{sec:results}. 

\subsubsection{Output metrics}
\label{meth:output_metrics}
In the 1D simulations, a non-uniform or `patterned' cell distribution arises in the form of cell clusters. Here, the number of cell clusters is considered to be the main outcome metric of interest. Additionally, cluster spacing and cluster width are related to characteristic pattern size in 2D, and can be observed \textit{in vitro} and qualitatively matched \cite{Tosin2006}. Hence, though not independent from the number of clusters, these are useful secondary metrics that can be used to assess biological plausibility of the model. In the following analyses, the number of cell clusters was calculated by applying a low threshold to the cell volume fraction distribution, and then identifying the number of discrete objects.

In 2D, cell clusters and/or vascular-like structures can arise. In this case, the spacing between clusters or the diameter of `voids' created by connected mesh-like patterning can be used to assess the spatial scale of patterning. To determine vessel width and spacing, several 1D slices of the 2D simulation output in each dimension are taken and treated in the same manner to the 1D model, using methods in \texttt{scipy.signal} to identify peaks and determine vessel widths. 

\subsubsection{Sensitivity analysis}
\label{meth:sa}
A sensitivity analysis can determine both the model tolerance to variation in parameters, \emph{i.e.} robustness, and the parameters which have the most influence over a particular model outcome. Sensitivity analyses can also play an important role in model selection by identifying parameters that have negligible impact on model outcomes, and are hence superfluous. It is beneficial to reduce the number of key parameters as more complex models require significantly more data than may be plausible to avoid over-fitting of the model. Additionally, this technique can be used to determine which changes in experimental variables have the most significant impact on vascular network formation in the model. 

In Python, an open source library \texttt{SALib} provides a sensitivity analysis framework for any user-developed function \cite{Herman2017}. \texttt{SALib} runs two functions: parameter space sampling (\texttt{SALib.sample}), and an analysis of the variation of model outputs with respect to the sampled parameter inputs (\texttt{SALib.analyze}). Here, the Sobol method was selected to provide the most detailed information for total, first-order, and second-order parameter interactions \cite{Saltelli2010}. The resulting Sensitivity indices (Si) are a measure, between 0 and 1, of a parameter's influence on the outcome metric. Very small values, comparative to the magnitude of the confidence level of the index, can be assumed to be negligible. 

\subsubsection{Particle swarm optimisation} 
\label{meth:pso}
Parameter optimisation is an umbrella term covering computational methods that seek to find an input parameter value, or set of input values, that produce an optimal model outcome as determined by a user-defined metric. Here, a particle swarm optimisation (PSO) algorithm has been selected which, given a metric to minimise, can find the optimal global or local point of the parameter set through an iterative procedure. In Python, the open source module \texttt{PySwarms} offers an implementation of this algorithm for any user-defined function and metric \cite{Miranda2018}.


\section{Results and discussion}
\label{sec:results}

\subsection{Effect of noisy initial conditions on cell cluster formation}
\label{res:init_cond_analysis}

To relate the initial conditions to the \textit{in vitro} scenario, a uniform distribution with some degree of random fluctuations on an appropriate spatial scale can be utilised. Implemented computationally, the three parameters related to this initial condition are: (i) the random seed; i.e. the specific number sequence used by the random number generator, (ii) the magnitude of the fluctuations, and (iii) the spatial scale of the random fluctuations, which refers to the length scale upon which the variability in cell volume fraction can occur, i.e. the width of a single cell or cluster of cells. 

A sensitivity analysis was performed on the three parameters that determine the initial condition. Table \ref{tab:SA_initialconditions} presents the total and first-order Sensitivity index (Si) for each of these, where 0 is given when the index is within the confidence interval. The Sensitivity index for magnitude is overall low, and for random seed and spatial scale is high. Considering first-order indices only, the random seed appears to have the most significant effect on outcome, suggesting that the high total Si for spatial scale is mostly due to second order interactions with the random seed. Given the high dependence of the number of cell clusters formed on the specific initial condition as characterised by the random seed, it is appropriate to treat the model stochastically, running analyses on model statistics over many random initial conditions, as opposed to individual outcomes. 

\begin{table}[h]
\setlength{\tabcolsep}{10pt} 
\renewcommand{\arraystretch}{1.5} 
\centering
\begin{tabular}{l l l l}
\hline
Parameter [units] & Value range & Total Si & First-order Si  \\ 
\hline
Seed [-] & 0 - 1000 & 0.92 & 0.22 \\
Spatial scale [$\mu$m] & 10 - 100 & 0.85 & 0 \\
Magnitude [\%] & 0 - 20 & 0.13 & 0 \\
\hline
\end{tabular}
\caption{Total and first-order sensitivity indices for parameters related to the cell volume fraction initial condition. The second column indicated the upper and lower bound for the parameter value used in the sensitivity analysis. The simulation results were analysed at $t$ = 4.}
\label{tab:SA_initialconditions}
\end{table}

\subsection{Effect of core model parameters on cell cluster formation}
\label{res:core_analysis}

Consider the core parameters $\alpha$, VEGF production rate, $\delta$, VEGF degradation rate, $\chi$, chemotaxis strength, and $\gamma_{nm}$, cell-matrix drag, identified by literature review to be most essential for the model. 

To verify that the parameter bounds informed by intuition and literature review are suitable for a sensitivity analysis, a manual parameter sweep was first conducted to gain some preliminary information about the regions of the parameter space that may lead to qualitatively distinct cell distribution outcomes. The dimensional parameter bounds used for the following analyses are given in Table \ref{tab:1Dcore_dim_param_bounds}.

\begin{table}[h]
\setlength{\tabcolsep}{10pt} 
\renewcommand{\arraystretch}{1.5} 
\centering
\begin{tabular}{l l l}
\hline
Core parameter [units] & Minimum value & Maximum value  \\ 
\hline
$\chi$ [Nm$^{-2}$] & 10 & 50 \\
$\gamma_{nm}$ [Nm$^{-4}s$] & 1 $\times$ 10$^{8}$ & 1 $\times$ 10$^{10}$ \\
$\alpha$ [g ml$^{-1}$s$^{-1}$] & $1 \times 10^{-11}$ & $1 \times 10^{-10}$ \\
$\delta$ [s$^{-1}$]  & $1 \times 10^{-4}$ & $1 \times 10^{-3}$ \\
\hline
\end{tabular}
\caption{Dimensional parameter bounds for the core model parameters used for computational model analysis.}
\label{tab:1Dcore_dim_param_bounds}
\end{table}

\begin{figure}[H]
    \centering
    \includegraphics[width=\textwidth]{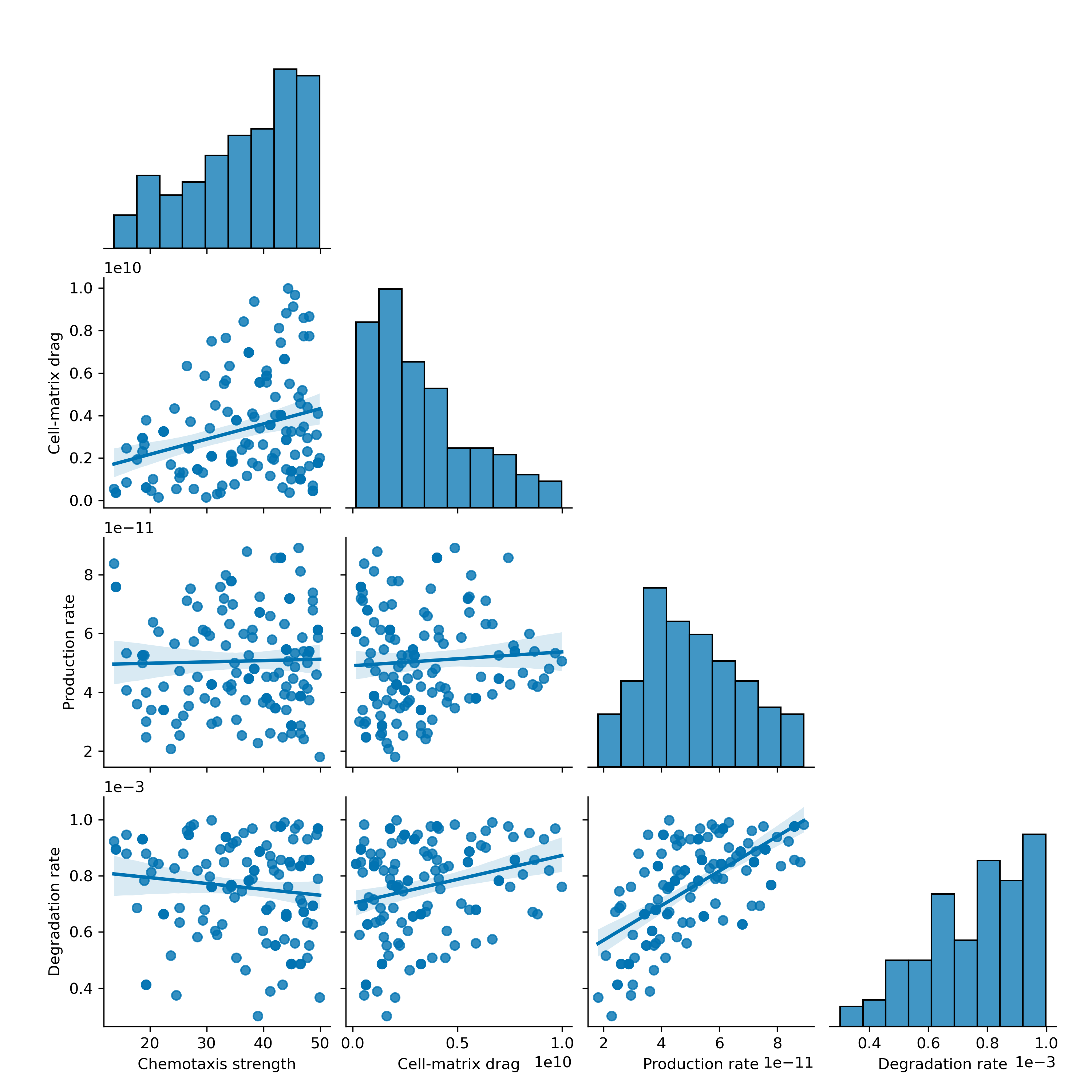}
    \caption{Pairplot illustration of the subset of parameter samples that were found to produce two or more clusters at $t$ = 2. The histograms present the distribution of the values of each parameter present in the subset; the off-diagonal scatter plots illustrate correlations between parameter values. A regression line is plotted for each off-diagonal scatter plot. Each quantity presented is dimensionless.}
    \label{fig:ps_pairplot}
\end{figure}

To conduct the parameter sweep, 768 parameter samples were selected using a Saltelli sampler, and the model was run for each sample until $t$ = 2 ($\sim$ 54 hours), recording the number of clusters observed. The profile of a subset of this data, for samples in which two or more clusters were observed at $t$ = 2, is presented in Figure \ref{fig:ps_pairplot}. The histograms presented on the diagonal illustrate the values of each parameter present in this subset, where a clear skew towards high or low values of each parameter can be identified for each of the core parameters. The off-diagonal scatter plots are used to illustrate potential relationships between two parameter values, with a regression line plotted for each. The off-diagonal plots with the clearest correlation between parameters include a positive relationship between VEGF production rate and degradation, and between chemotaxis strength and cell-matrix drag. 

\begin{figure}[H]
    \centering
    \begin{subfigure}[b]{0.49\textwidth}
    \centering
    \includegraphics[width=\textwidth]{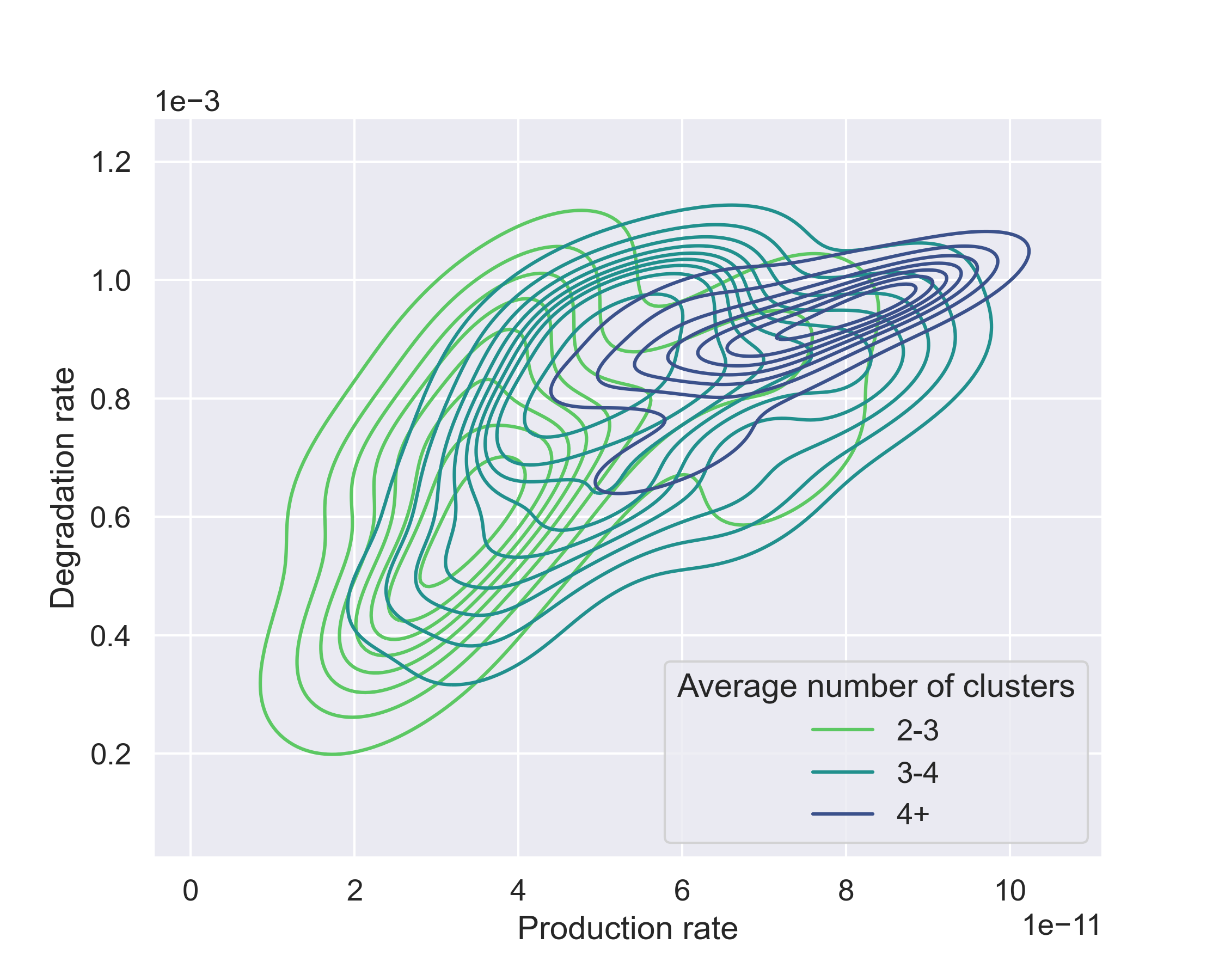}
    \caption{VEGF production against VEGF degradation}
    \label{fig:ps_prod_deg_kde}
    \end{subfigure}
    \centering
    \begin{subfigure}[b]{0.49\textwidth}
    \centering
    \includegraphics[width=\textwidth]{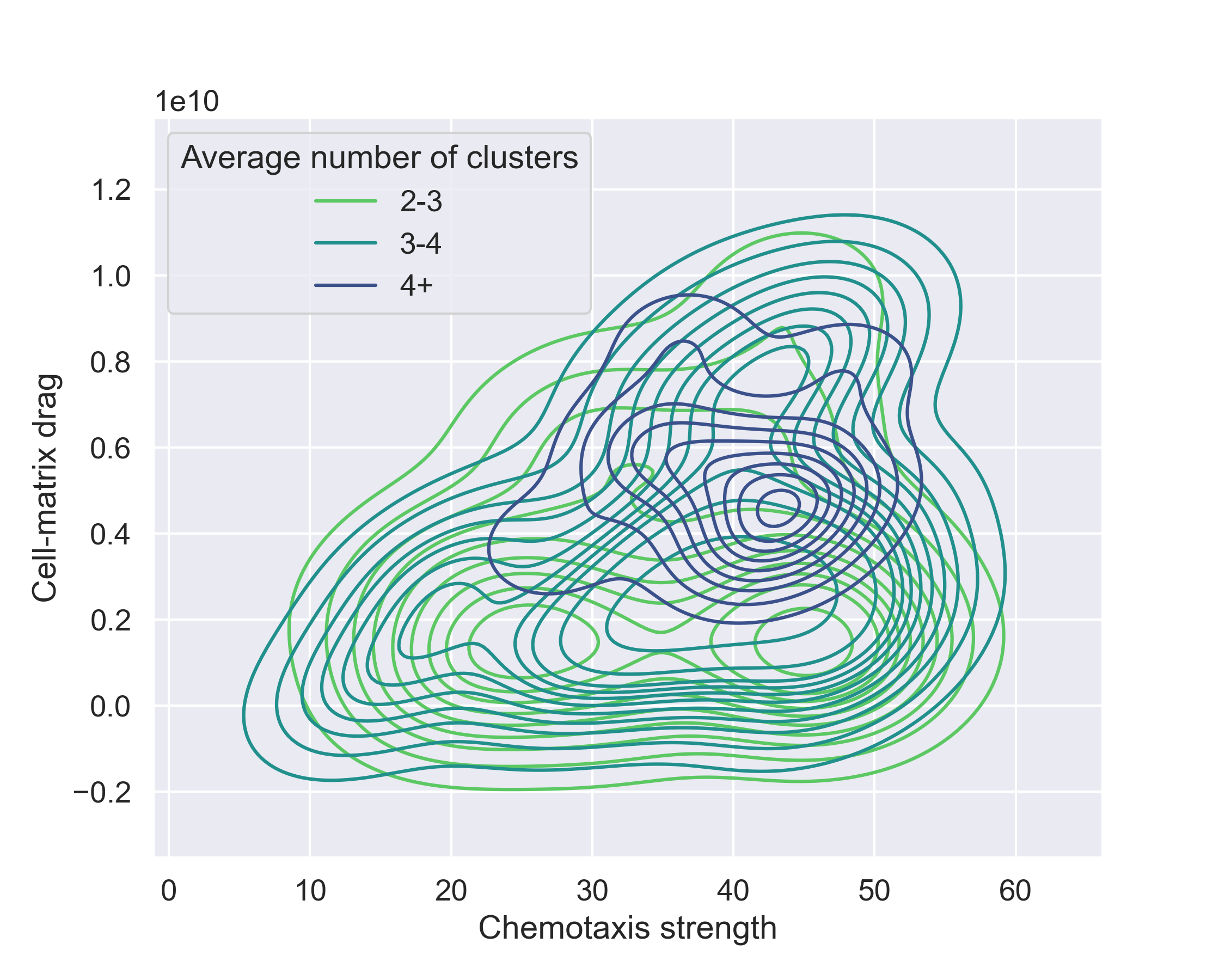}
    \caption{Chemotaxis strength against cell-matrix drag}
    \label{fig:ps_chem_drag_pde}
    \end{subfigure}
    
    \caption{Kernel density estimate (KDE) plots illustrating the relationship between dimensionless model parameters based on a parameter sweep. The KDE is plotted for the data subset that produced 2 - 3, 3 - 4, and more than 4 clusters respectively.} 
    \label{fig:ps_kdeplots}
\end{figure}

In Figure \ref{fig:ps_kdeplots}, the parameter samples are labelled based on the specific number of clusters produced. Figure \ref{fig:ps_prod_deg_kde} illustrates the values of VEGF production rate and VEGF degradation rate for which either 2 - 3, 3 - 4, or more than 4 clusters were produced. Likewise,  Figure \ref{fig:ps_chem_drag_pde} illustrates the values of chemotaxis strength and cell-matrix drag which produce different numbers of clusters. In both, a positive relationship between parameters is clear, and high values of each parameter are required to achieve 4 or more clusters.

A sensitivity analysis was also conducted on these four parameters, using the number of clusters formed as the output metric. The analysis was based on the average simulation output over 10 random initial conditions. The same 10 random seeds are used in the computational analysis throughout this paper. Table \ref{tab:sa_coreparameters} presents the total Sensitivity index for each parameter at $t$ = 1, 2, 3 and 4 respectively. The decrease in Si over time for chemotaxis strength and cell-matrix drag suggests that these parameters play a role in the timing of cluster formation, but are less influential over the number of clusters eventually formed. The most influential parameters on cell clustering hence relate to VEGF production and degradation. 

\begin{table}[h]
\setlength{\tabcolsep}{10pt} 
\renewcommand{\arraystretch}{1.5} 
\centering
\begin{tabular}{l l l l l}
\hline
$t$ [-] & 1 & 2 & 3 & 4  \\
\hline
VEGF degradation rate & 0.67 & 0.76 & 0.81 & 0.84 \\
VEGF production rate & 0.55 & 0.58 & 0.59 & 0.58 \\
Chemotaxis strength & 0.40 & 0.35 & 0.31 & 0.29 \\
Cell-matrix drag & 0.34 & 0.25 & 0.23 & 0.2 \\
\hline

\hline
\end{tabular}
\caption{Total Sensitivity index for each of the core parameters, based on number of cell clusters formed, at $t$ = 1, 2, 3, and 4 respectively. The parameter bounds used are given in Table \ref{tab:1Dcore_dim_param_bounds}.}
\label{tab:sa_coreparameters}
\end{table}

\subsection{Minimum spatial scale of core model cell cluster formation}
\label{res:min_spac_clust}

In the parameter sweep presented above, the maximum average number of clusters obtained across the 10 random initial conditions is 4.7, suggesting that it is uncommon for 5  or more clusters to be produced by the core four-parameter model within the parameter bounds given. To verify the maximum number of clusters the core model is able to produce, and for which parameter sample, a particle swarm optimisation (PSO) algorithm was utilised to maximise the average number of clusters produced within the parameter bounds of Table \ref{tab:1Dcore_dim_param_bounds}. 

The optimal parameter values found by PSO to maximise the average number of clusters at $t$ = 2 and $t$ = 4 are presented in Table \ref{tab:po_1D_core}. These indicate that generally 5-6 clusters can be achieved with relatively high rates of VEGF production and degradation, given the prescribed limits. 

\begin{table}[H]
\setlength{\tabcolsep}{10pt} 
\renewcommand{\arraystretch}{1.5} 
\centering
\begin{tabular}{l l l l l}
\hline
$t$ [-] & 2 & 4 \\
\hline
average no. of clusters & $5.0$ & $5.4$ \\ 
VEGF degradation rate & $9.96 \times 10^{-4}$ & $9.0 \times 10^{-4}$ \\
VEGF production rate & $7.41 \times 10^{-11}$ & $8.92 \times 10^{-11}$ \\
Chemotaxis strength & $36.7$ & $40.1$ \\
Cell-matrix drag & $4.62 \times 10^9$ & $4.34 \times 10^9$ \\
\hline
\hline
\end{tabular}
\caption{Dimensionless parameter values found to maximise the number of cell clusters formed at $t$ = 2 and $t$ = 4 respectively, using the particle swarm optimisation method.}
\label{tab:po_1D_core}
\end{table}

\begin{figure}[H]
\hfill
\begin{subfigure}{0.46\textwidth}
    \centering
    \includegraphics[width = \textwidth]{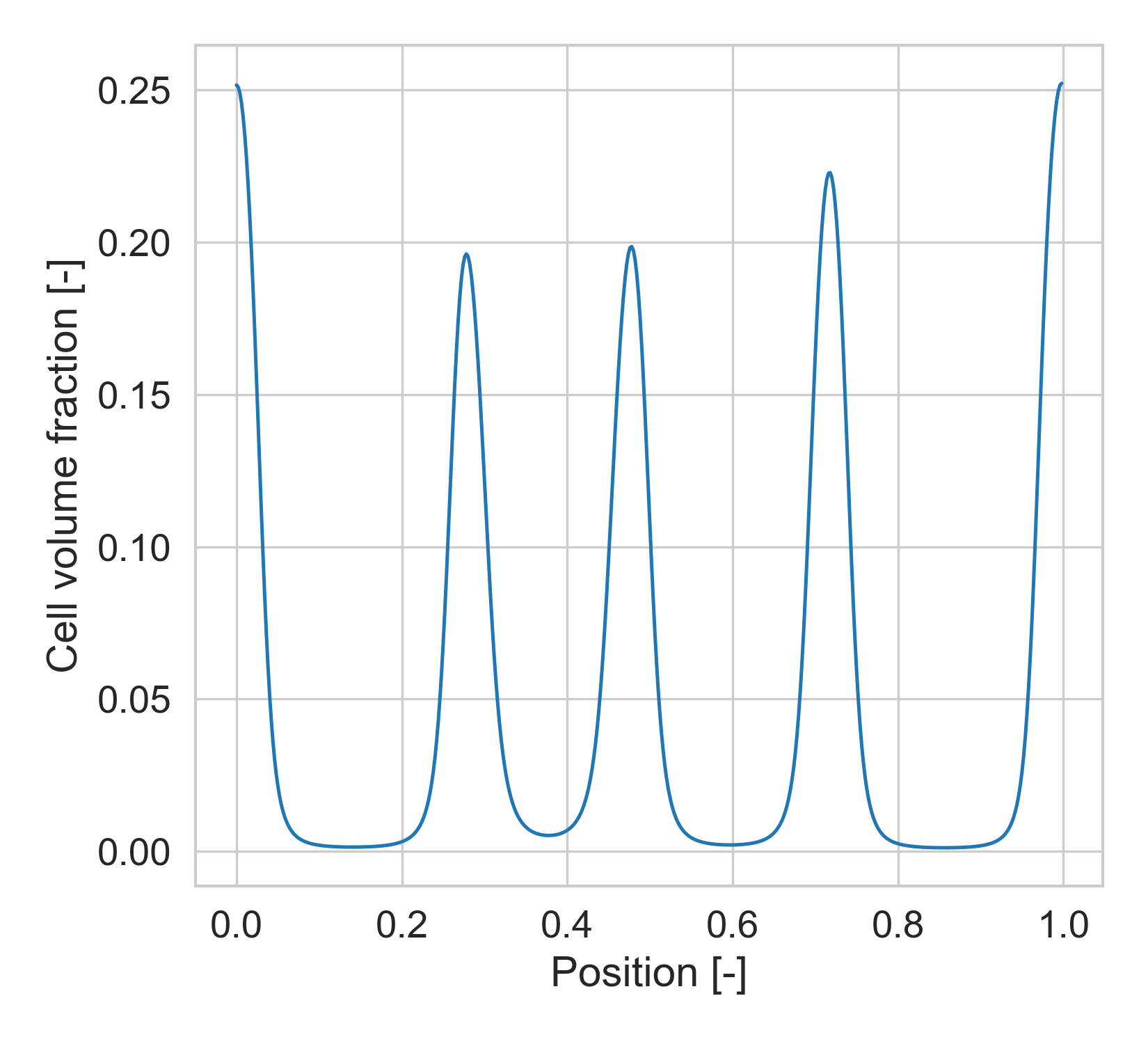}
    \caption{Example cell cluster formation in 1D}
    \label{fig:po_1D_example}
\end{subfigure}
\hfill
\begin{subfigure}{0.52\textwidth}
    \centering
    \includegraphics[width = \textwidth]{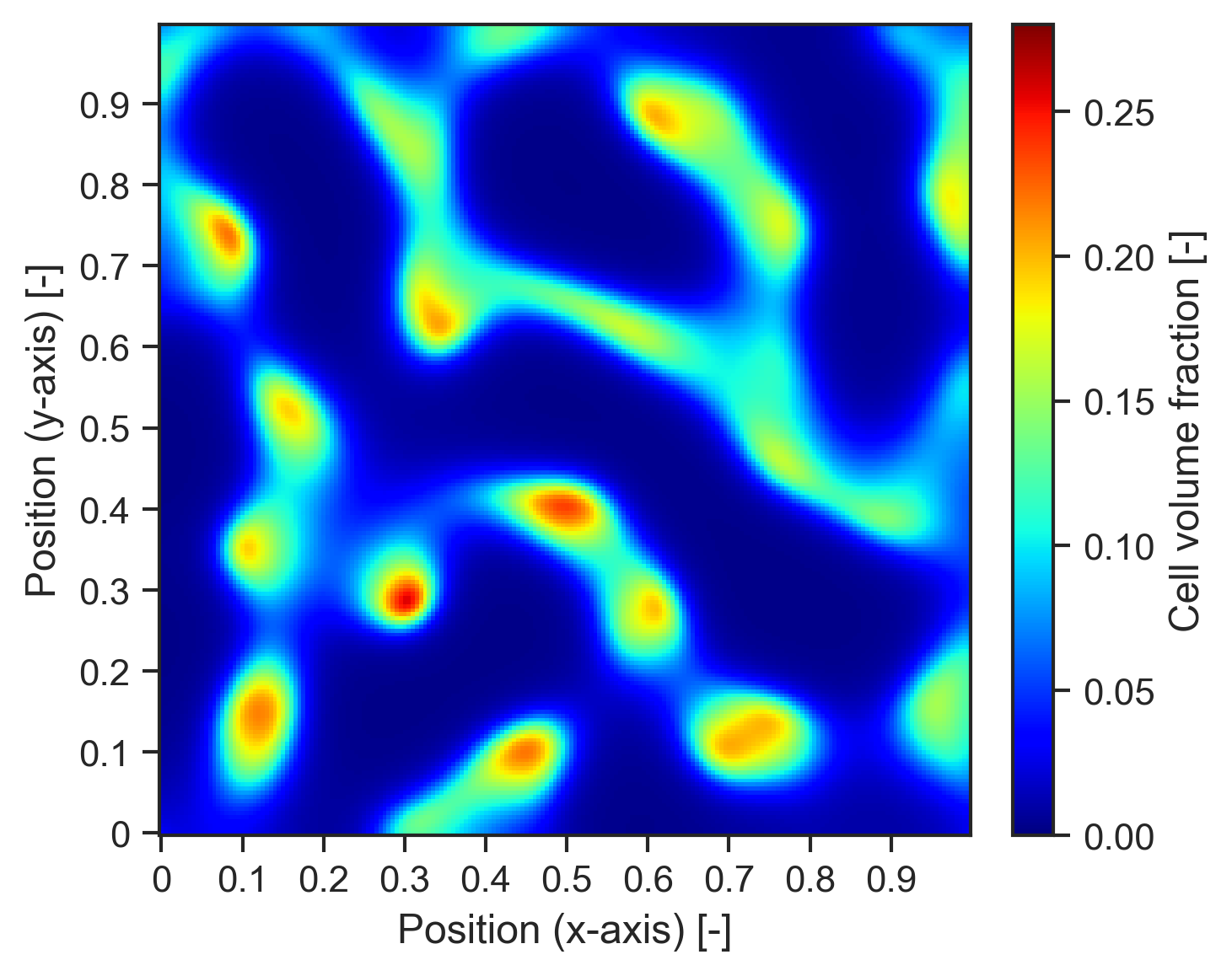}
    \caption{Example cell pattern formation in 2D}
    \label{fig:po_2D_example}
\end{subfigure}
\caption{Example cell cluster and pattern formation for the optimised parameter values given in the first column ($t$ = 2) of Table \ref{tab:po_1D_core}, shown at $t$ = 2. The same random initial condition was used to generate all 1D and 2D example plots.}
\end{figure}

Figure \ref{fig:po_1D_example} illustrates an example of cell clustering produced by this core model, using the optimised parameter values given in Table \ref{tab:po_1D_core}, in which 5 clusters are present at $t$ = 2. Assuming the width of each cluster and spacing between the clusters is the same, this suggests a typical cluster or spacing width of $\sim$ 350 $\mu$m. This is larger than the typical spacing seen between vascular components \textit{in vitro} and \textit{in vivo} \cite{Serini2003}. Figure \ref{fig:po_2D_example} presents an example of the 2D model output also at $t$ = 2 using these parameter values. The spatial scale of the 2D cell structures is similar to the 1D clusters, and this is explored further in Section \ref{res:spac_anal_all}. 

\subsection{Long-term behaviour of cell clusters}
\label{res:long_clust}

Based on the optimised parameter values given in Table \ref{tab:po_1D_core}, Table \ref{tab:po_1D_stability} shows the average number of clusters seen at longer time points up to and including $t$ = 20. Though the model is currently intended to model initial pattern formation, not long-term cell behaviour, Table \ref{tab:po_1D_stability} demonstrates that patterning driven by autologous chemotaxis in this model can support long-term cell cluster formation. 

\begin{table}[H]
\setlength{\tabcolsep}{10pt} 
\renewcommand{\arraystretch}{1.5} 
\centering
\begin{tabular}{l l l l l}
\hline
$t$ [-] & 2 & 4 & 10 & 20 \\
\hline
No. of clusters & 5.9 & 5.5 & 5.1 & 4.6 \\
\hline
\hline
\end{tabular}
\caption{Persistence of total number of clusters averaged over 10 initial conditions. Parameter values used were taken from Table \ref{tab:po_1D_core} ($t$ = 4 column).}
\label{tab:po_1D_stability}
\end{table}

\begin{figure}[H]
    \hfill
    \begin{subfigure}[b]{0.49\textwidth}
    \centering
    \includegraphics[width=\textwidth]{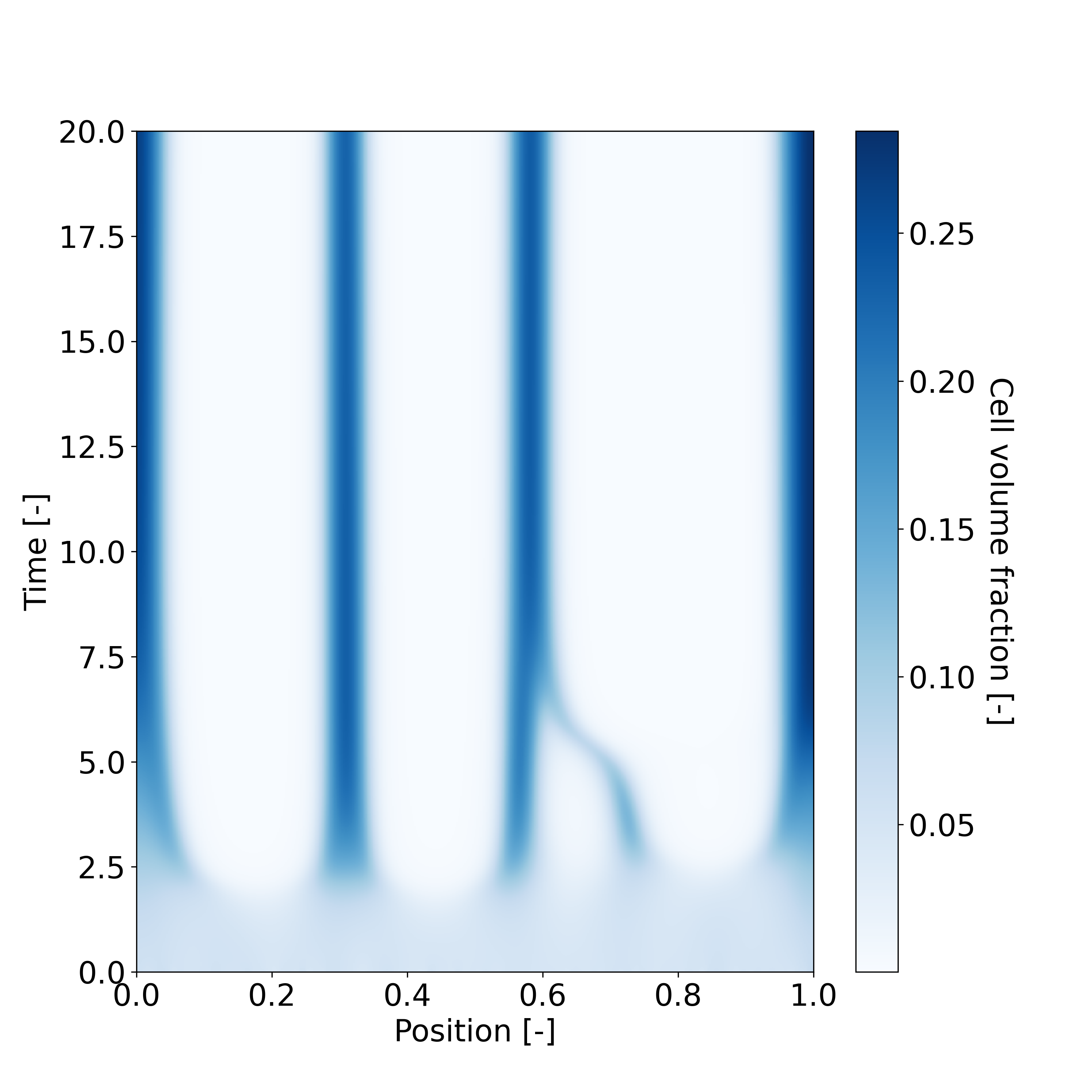}
    \caption{Example 1}
    \label{fig:clust_pers_1}
    \end{subfigure}
    \hfill
    \begin{subfigure}[b]{0.49\textwidth}
    \centering
    \includegraphics[width=\textwidth]{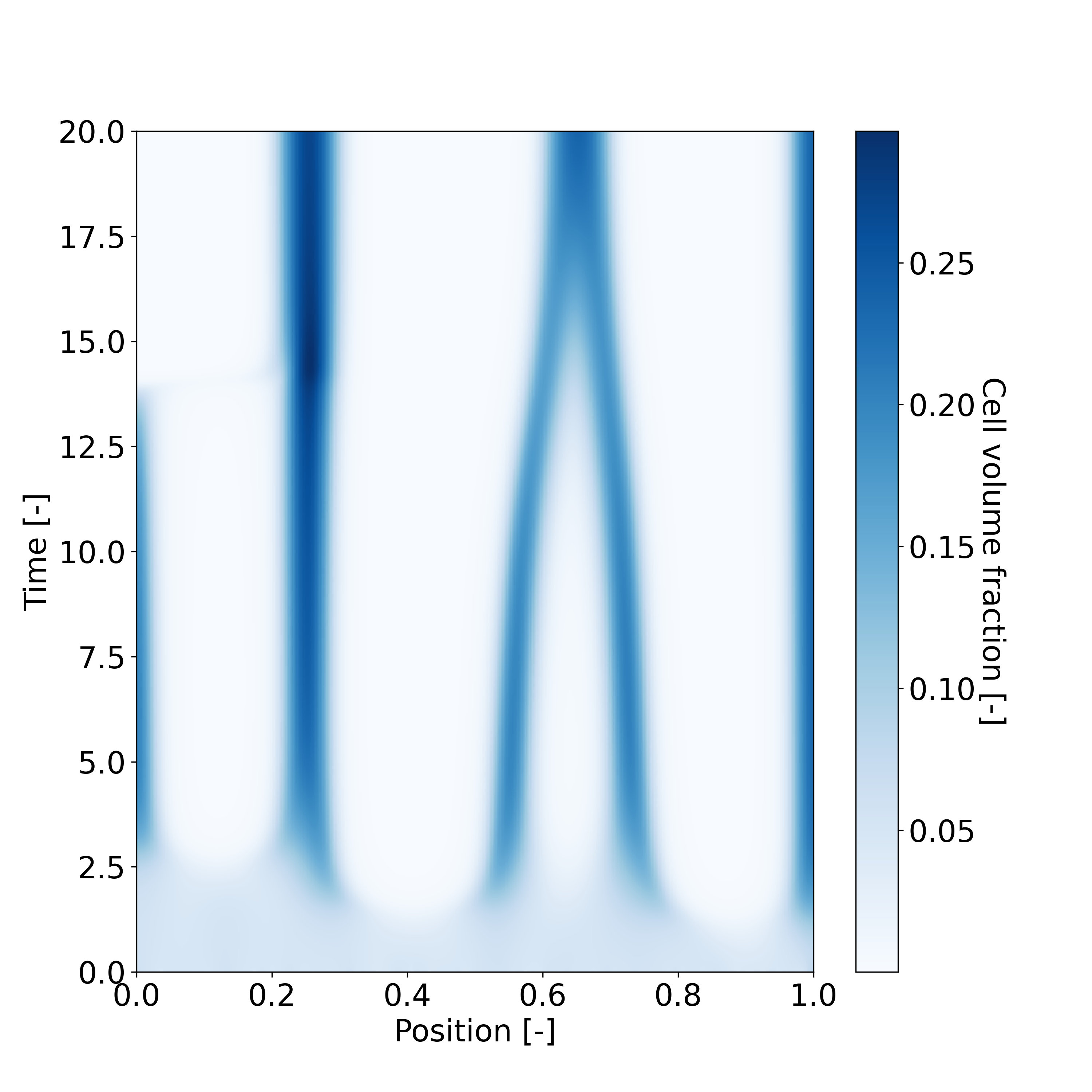}
    \caption{Example 2}
    \label{fig:clust_pers_2}
    \end{subfigure}
    
    \hfill
    \begin{subfigure}[b]{0.49\textwidth}
    \centering
    \includegraphics[width=\textwidth]{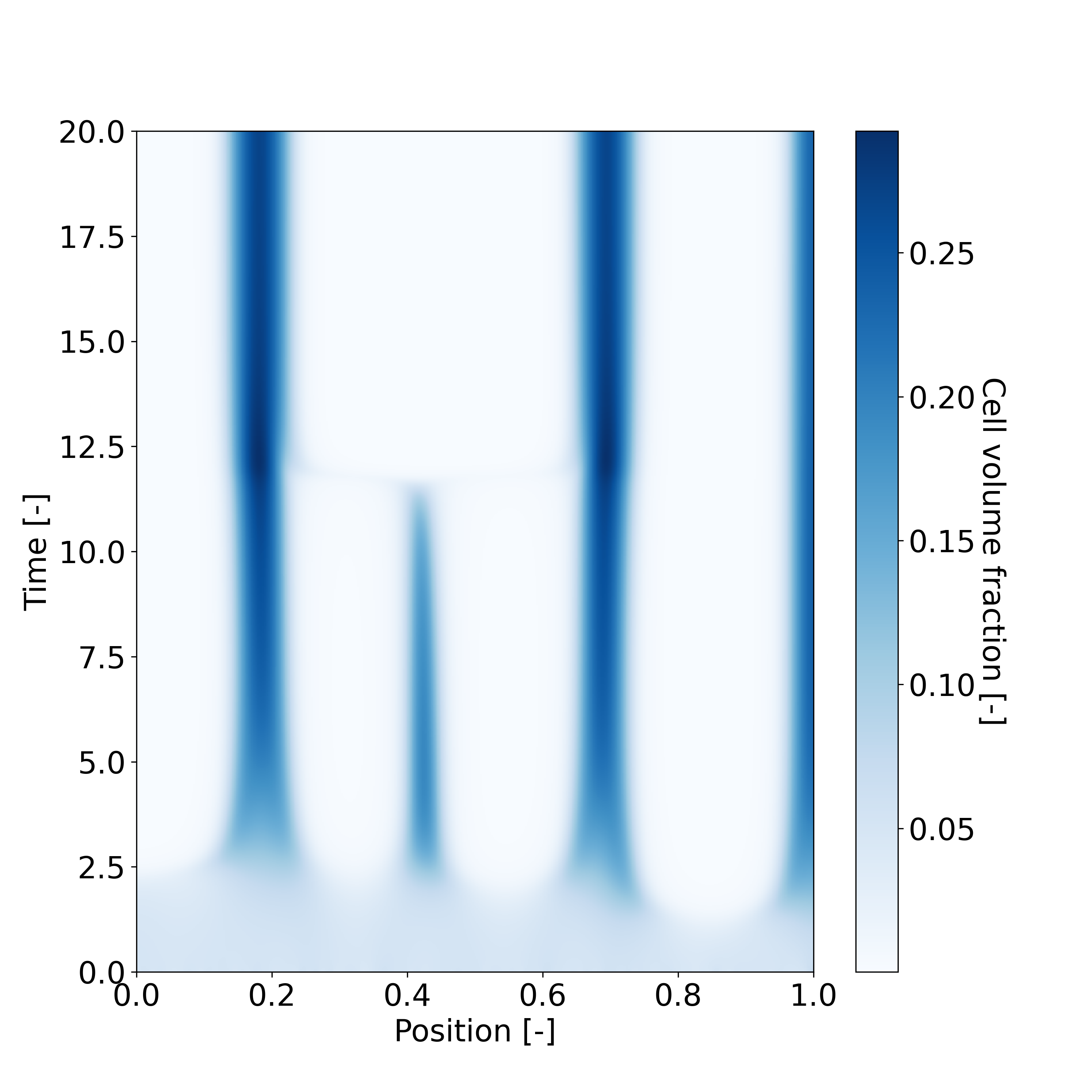}
    \caption{Example 3}
    \label{fig:clust_pers_3}
    \end{subfigure}
    \hfill
    \begin{subfigure}[b]{0.49\textwidth}
    \centering
    \includegraphics[width=\textwidth]{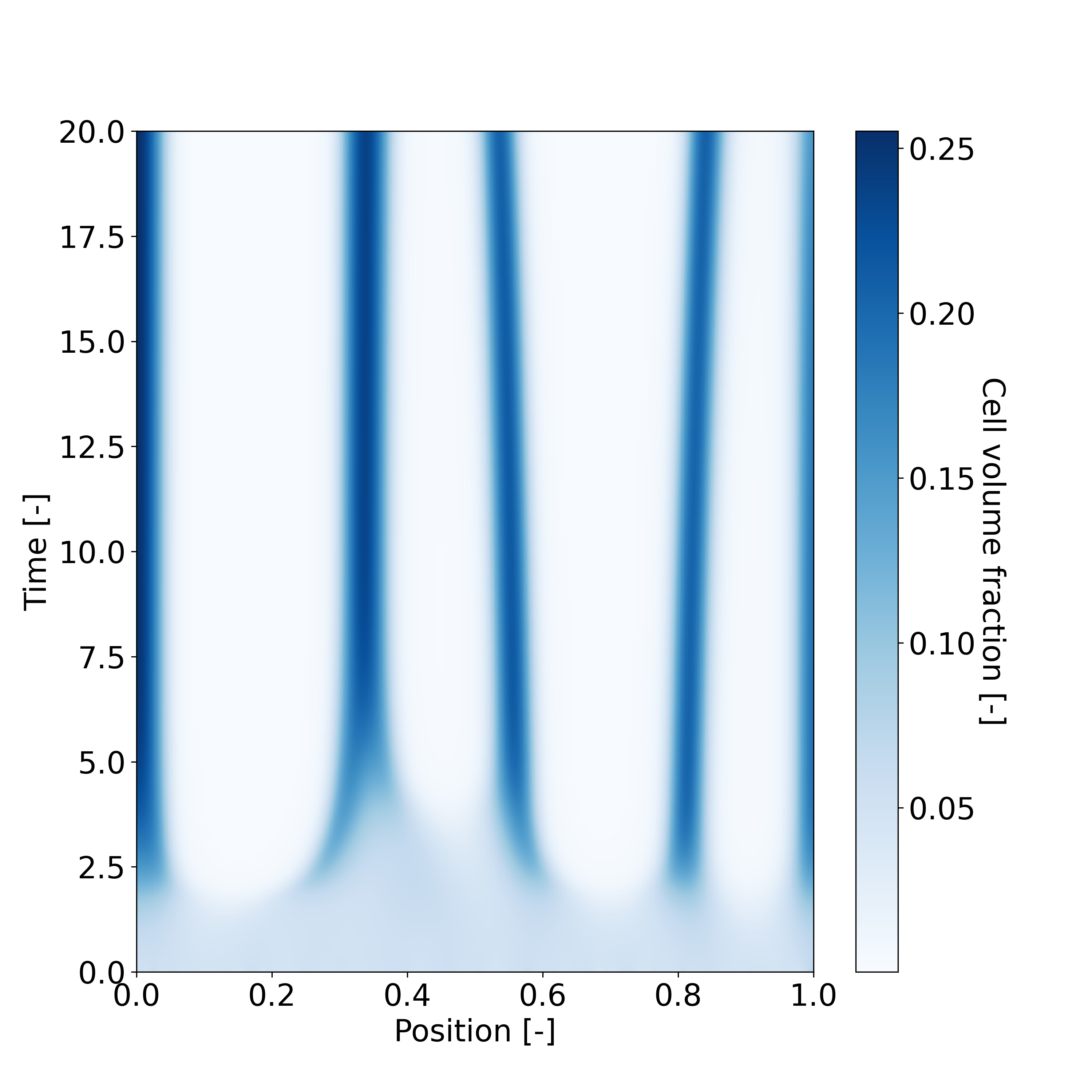}
    \caption{Example 4}
    \label{fig:clust_pers_4}
    \end{subfigure}
    
    \caption{Examples of cell cluster persistence and cluster merging over time, using optimised parameter values given in the second column ($t$ = 4) of Table \ref{tab:po_1D_core}. Four initial conditions were chosen randomly to reflect the range of long-term behaviours observed.} 
    \label{fig:clust_pers_core}
\end{figure}

To illustrate the long-term dynamics of the cell clusters, Figure \ref{fig:clust_pers_core} presents four examples of cluster persistence, merging, and dissipation that can take place, based on four different random initial conditions. Each example shows the cell volume fraction along the 1D geometry, with time on the y-axis up to $t$ = 20. In Figure \ref{fig:clust_pers_1}, 5 clusters are observed to form between $t$ = 2.5 and $t$ = 5, before a cluster merging event takes place around $t$ = 5 to $t$ = 7.5. After $t$ = 7.5, four clusters persist appearing stable. Additional analysis is required to conclude whether the system in this case has reached a steady state. Similar merging events take place in Figures \ref{fig:clust_pers_2} and \ref{fig:clust_pers_3}. In Figure \ref{fig:clust_pers_4}, five clusters persist beyond $t$ = 20, however there is some displacement of the clusters' positions, indicating that a steady state has not yet been obtained.  

\subsection{Impact of additional mechanisms on cell cluster formation}
\label{res:add_mech}

Given the limits of the core model investigated, in which the minimum spatial scale of patterning was found to be $\sim$ 350 $\mu$m, some common additional mechanisms are now considered to determine their potential impact on this patterning scale. These include: (i) VEGF uptake, (ii) cell-matrix traction, (iii) cell-cell aggregation, and (iv) cell-cell contact inhibition.

The first additional mechanism, VEGF uptake, is characterised by two parameters (see Section \ref{model:vegf_reaction}): uptake rate, $\kappa$, and the VEGF concentration at which uptake is half-maximal, $K$. The second, cell-matrix traction, is an additional interphase force in Equations (\ref{eqn:2Dun}) and (\ref{eqn:2Dvn}), characterised by the traction force parameter $\eta$. The final two, aggregation and contact inhibition, are included as cell intraphase pressures in the additional pressure term $\Pi$ shown in the general Equation~(\ref{simpmomn}), and expanded in Equations (\ref{eqn:2Dun}) and (\ref{eqn:2Dvn}). These two mechanisms are characterised by the parameters $\nu$, an aggregation pressure, and $\delta_n$, a contact inhibition pressure. 

To determine the impact that each of these parameters has on the spatial scale of pattern formation in the model, the number of cell clusters produced under the same core parameter regime and initial conditions was determined for a range of values of these additional parameters. The dimensional parameter values considered for each of the assessed parameters are given in Table \ref{tab:1Dadd_dim_param_bounds}. 

\begin{figure}[p!]
    \hfill
    \begin{subfigure}[b]{0.45\textwidth}
    \centering
    \includegraphics[width=\textwidth]{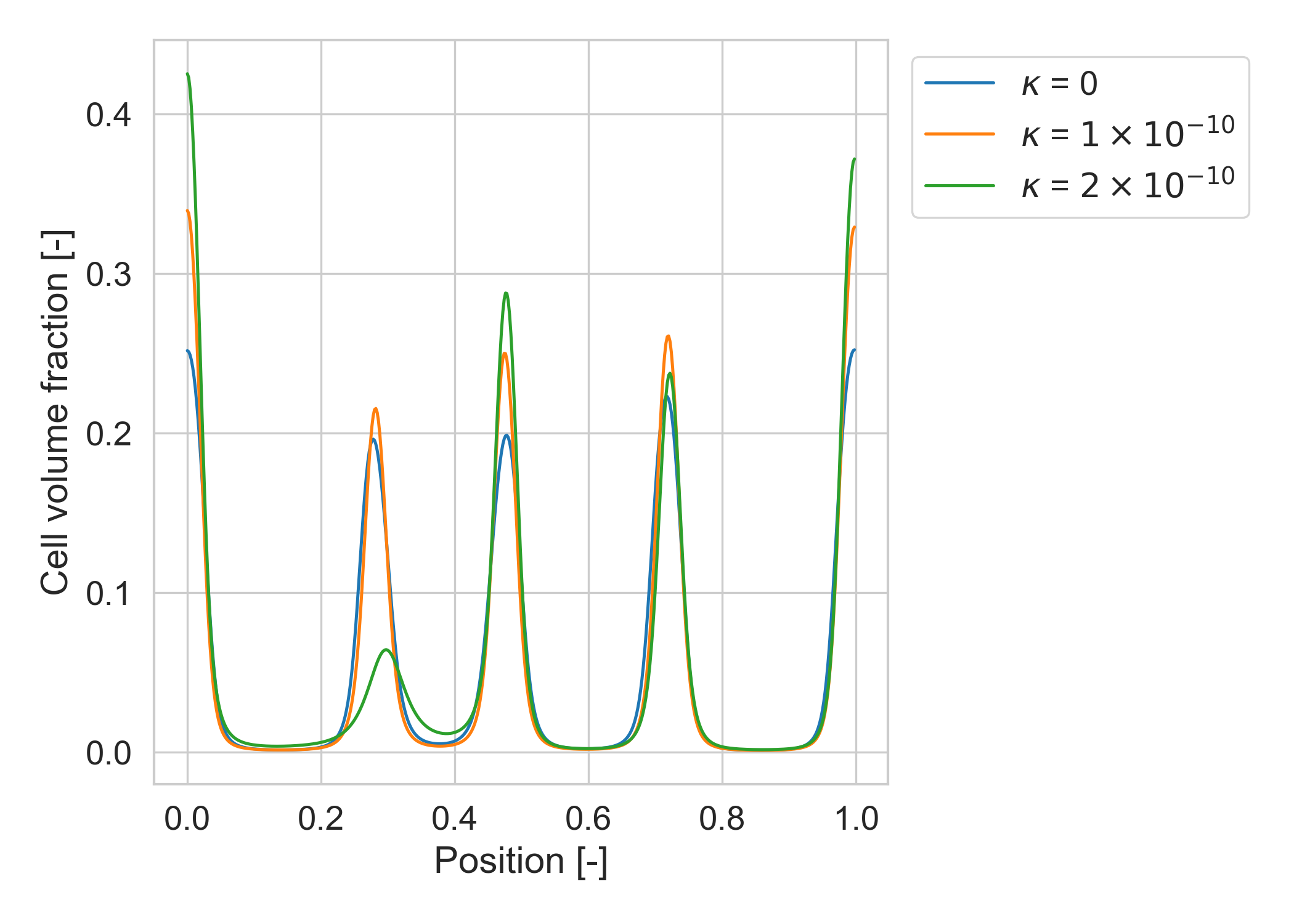}
    \caption{VEGF uptake rate, $\kappa$}
    \label{fig:add_uptake_1D}
    \end{subfigure}
    \hfill
    \begin{subfigure}[b]{0.42\textwidth}
    \centering
    \includegraphics[width=\textwidth]{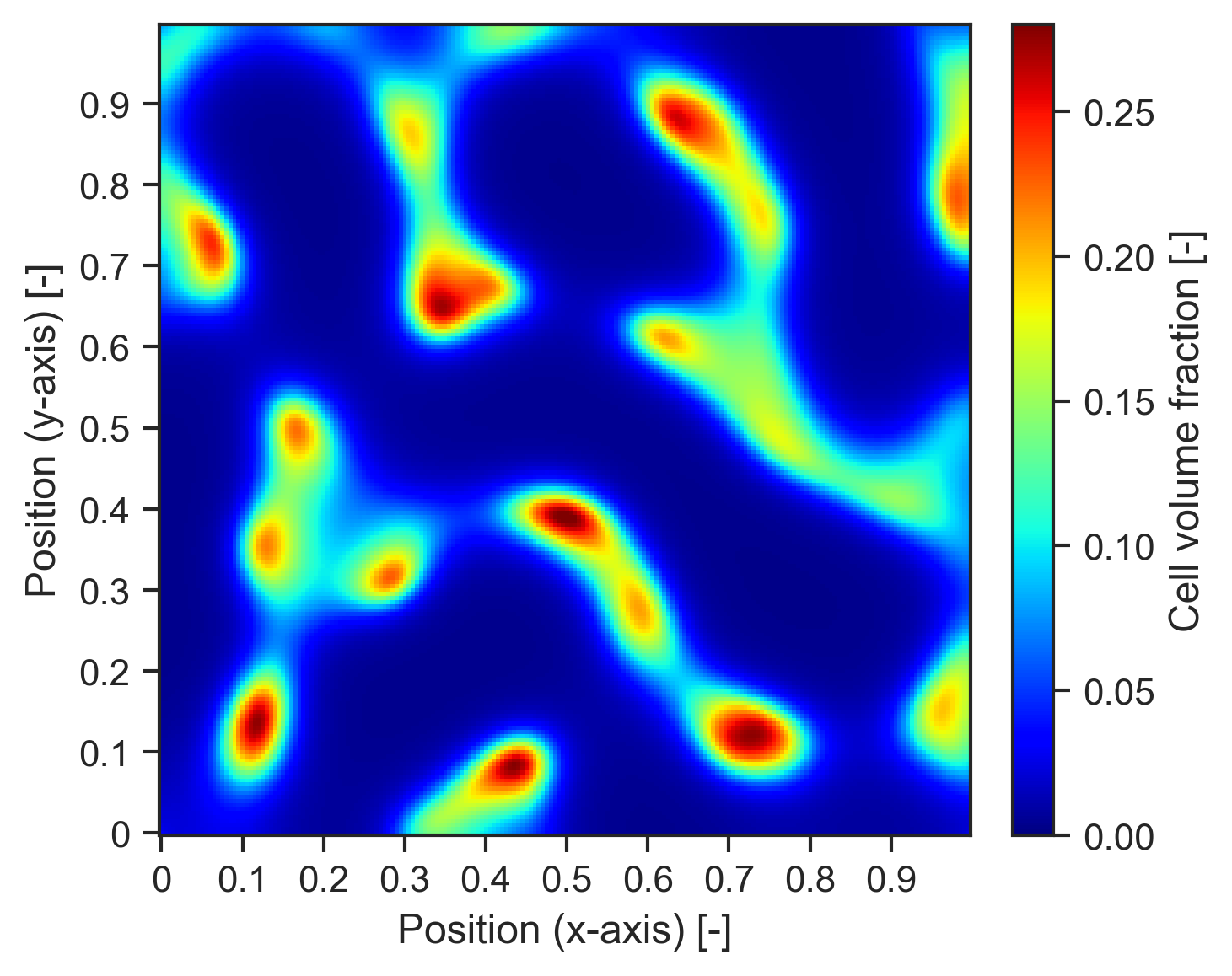}
    \caption{VEGF uptake rate, $\kappa$ = 1 $\times$ 10$^{-10}$}
    \label{fig:add_uptake_2D}
    \end{subfigure}
    
    \hfill
    \begin{subfigure}[b]{0.45\textwidth}
    \centering
    \includegraphics[width=\textwidth]{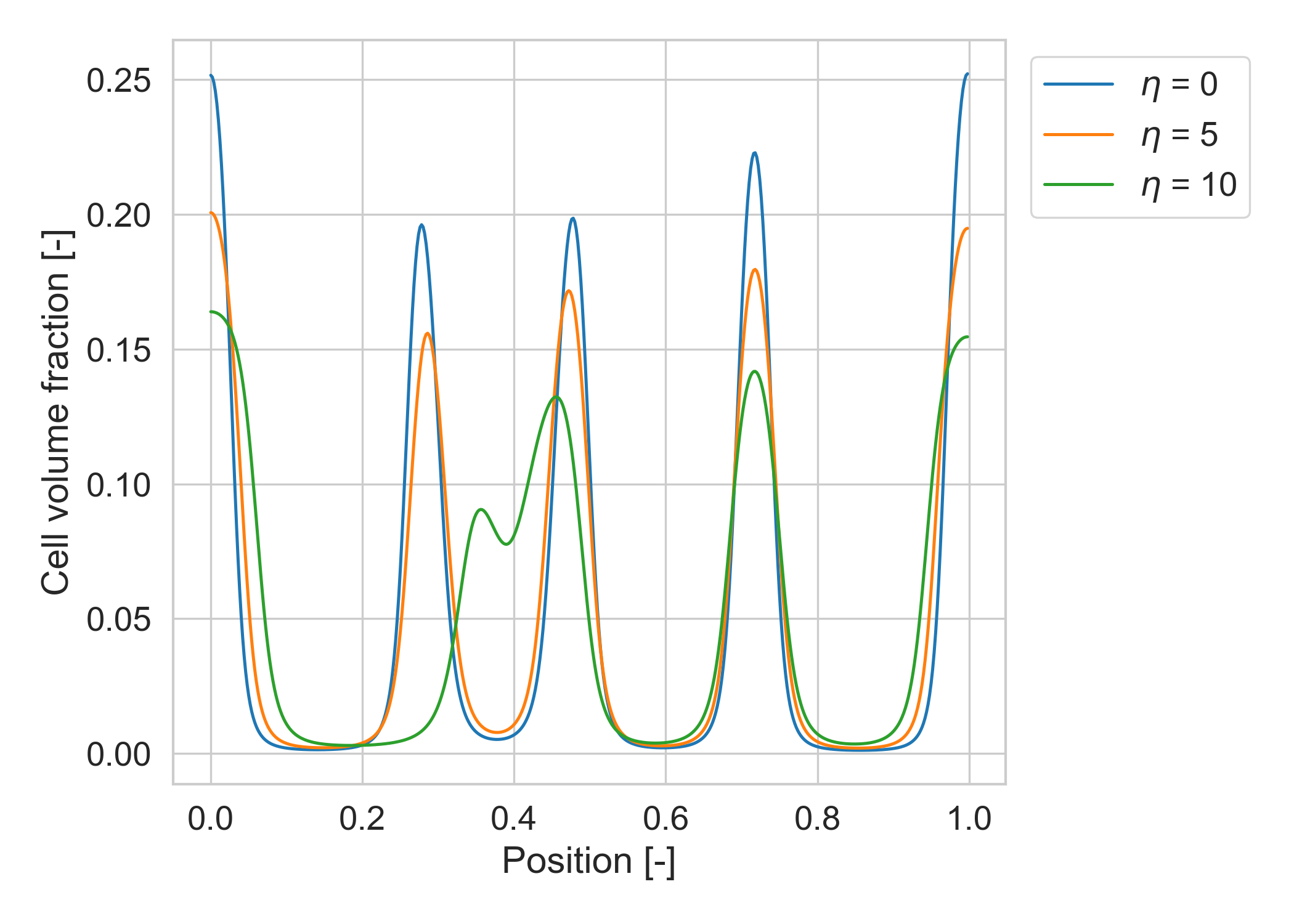}
    \caption{Cell-matrix traction, $\eta$}
    \label{fig:add_trac_1D}
    \end{subfigure}
    \hfill
    \begin{subfigure}[b]{0.42\textwidth}
    \centering
    \includegraphics[width=\textwidth]{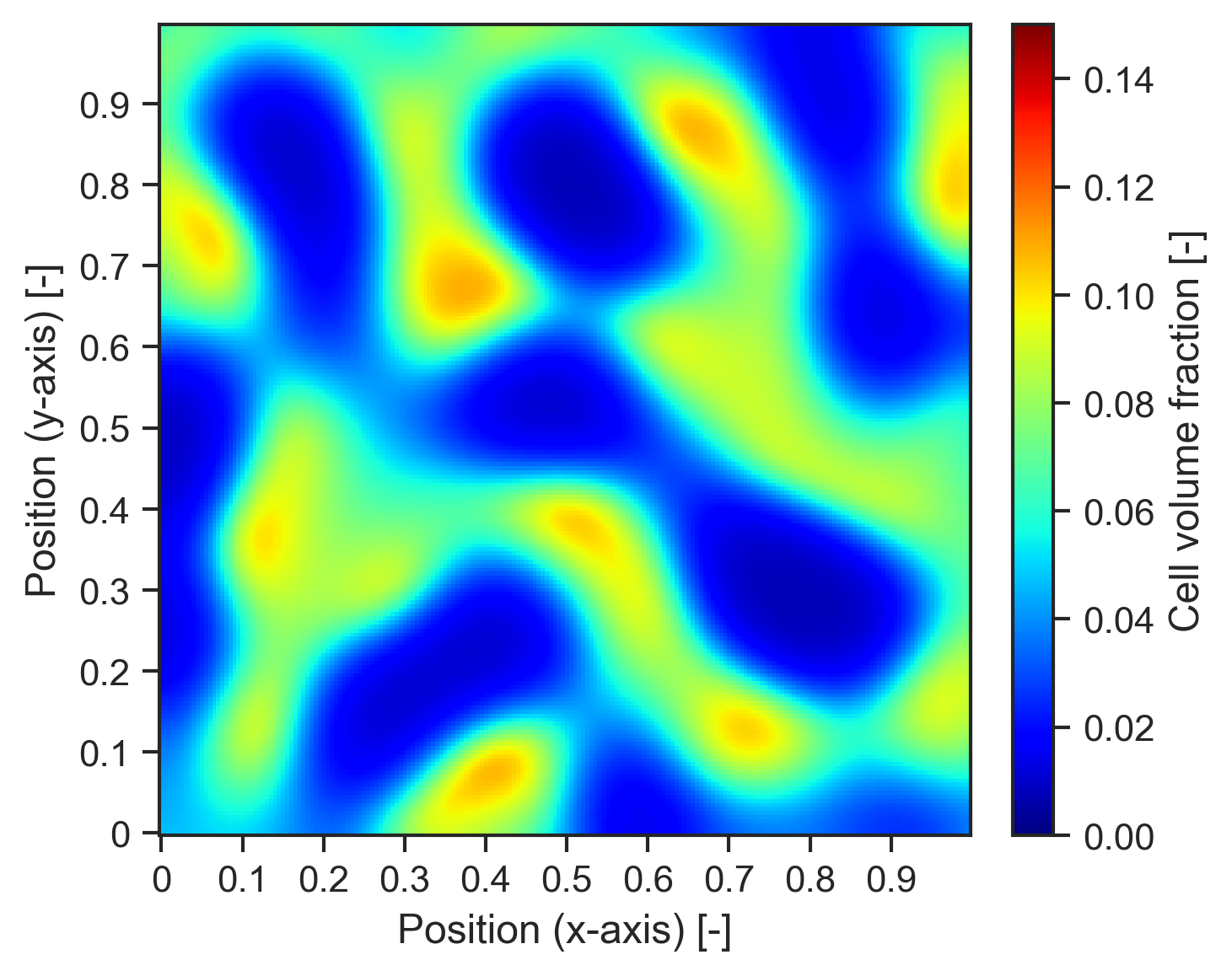}
    \caption{Cell-matrix traction, $\eta = 10$}
    \label{fig:add_trac_2D}
    \end{subfigure}
    
    \hfill
    \begin{subfigure}[b]{0.45\textwidth}
    \centering
    \includegraphics[width=\textwidth]{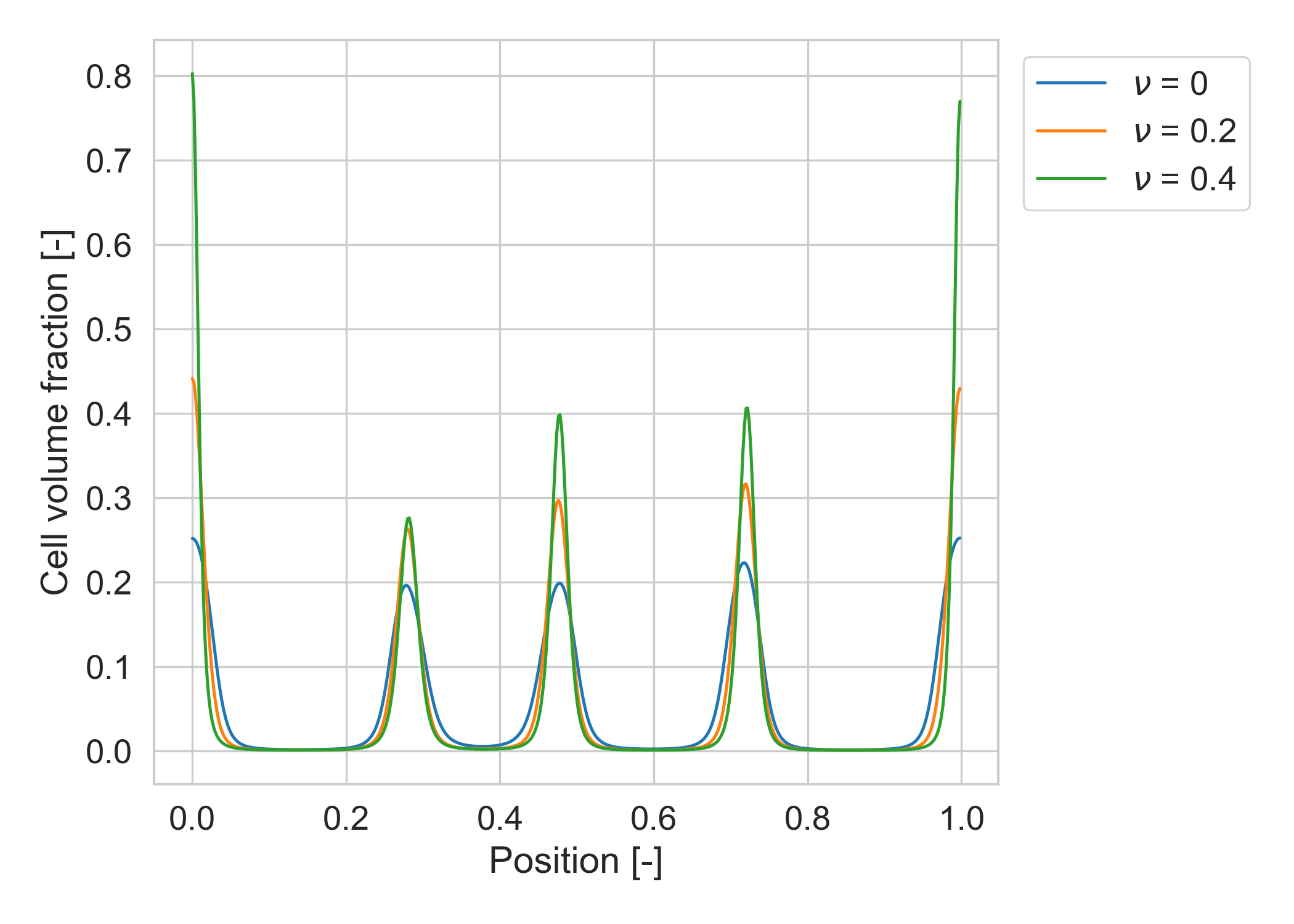}
    \caption{Cell-cell aggregation, $\nu$}
    \label{fig:add_agg_1D}
    \end{subfigure}
    \hfill
    \begin{subfigure}[b]{0.42\textwidth}
    \centering
    \includegraphics[width=\textwidth]{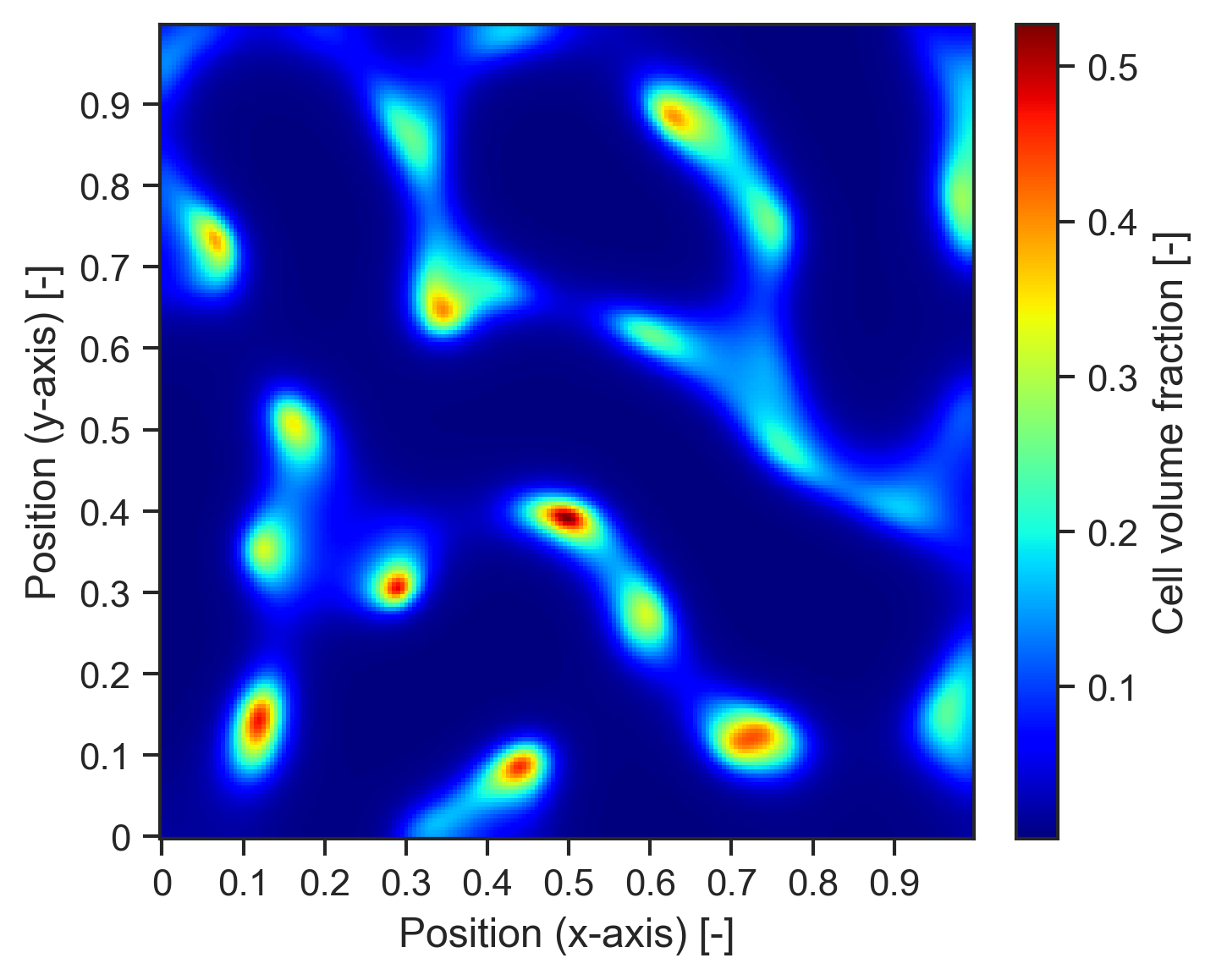}
    \caption{Cell-cell aggregation, $\nu = 0.5$}
    \label{fig:add_agg_2D}
    \end{subfigure}
    
    \hfill
    \begin{subfigure}[b]{0.45\textwidth}
    \centering
    \includegraphics[width=\textwidth]{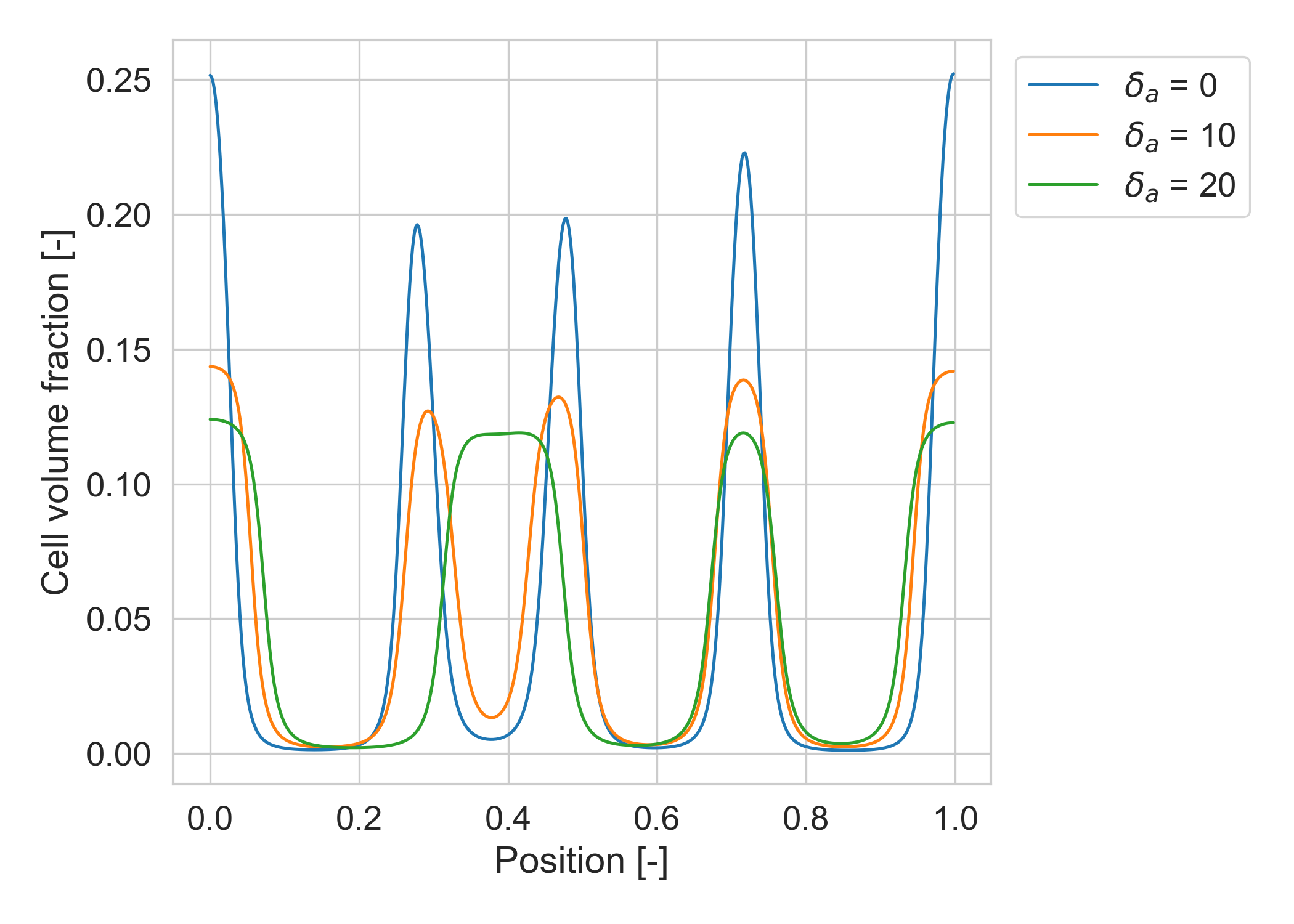}
    \caption{Contact inhibition, $\delta_a$}
    \label{fig:add_inhib_1D}
    \end{subfigure}
    \hfill
    \begin{subfigure}[b]{0.42\textwidth}
    \centering
    \includegraphics[width=\textwidth]{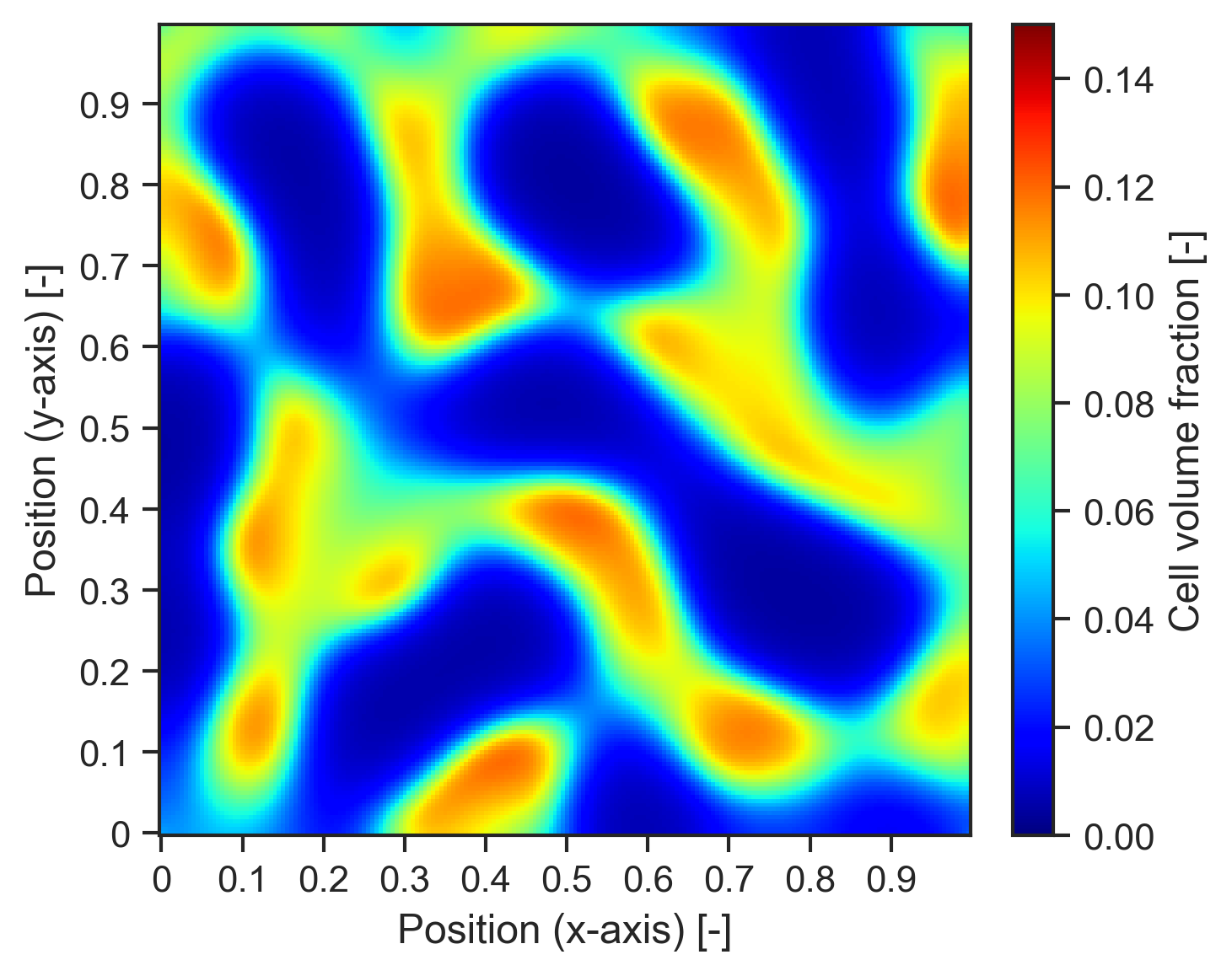}
    \caption{Contact inhibition, $\delta_a = 10$}
    \label{fig:add_inhib_2D}
    \end{subfigure}

    \caption{Examples of the impact of additional parameters in 1D and 2D, each shown at $t$ = 2. The core parameter values used were fixed based on Table \ref{tab:po_1D_core} ($t$ = 2 column), and the same initial condition was used to generate each example.} 
    \label{fig:add_params_ex}
\end{figure}

Table \ref{tab:1Dadd_dim_results} presents the average number of cell clusters formed, averaged over the same 10 initial conditions as used previously, for each additional parameter. It is clear from these results that, although there is a clear impact of these parameters on the outcome in terms of the number of clusters formed, they are unable to increase the number of clusters and hence reduce the spatial scale of patterning observed. Figure \ref{fig:add_params_ex} further illustrates the impact of each of these parameters, using a fixed initial condition and fixed core parameter values, in both 1D and 2D.


\begin{table}[H]
\setlength{\tabcolsep}{10pt} 
\renewcommand{\arraystretch}{1.5} 
\centering
\begin{tabular}{l l l}
\hline
Parameter [units] & First value & Second value  \\ 
\hline
$\kappa$ [g ml$^{-1}$s$^{-1}$] & $2 \times 10^{-10}$ & $5 \times 10^{-10}$ \\
$\eta$ [Nm$^{-2}$] & $10$ & $20$ \\
$\nu$ [Nm$^{-2}$] & $0.3$ & $0.6$ \\
$\delta_n$ [Nm$^{-2}$] & $10$ & $20$ \\
\hline
\end{tabular}
\caption{Dimensional parameter values of the additional model parameters used to determine which additional parameters may play a role in cell clustering.}
\label{tab:1Dadd_dim_param_bounds}
\end{table}

\begin{table}[H]
\setlength{\tabcolsep}{10pt} 
\renewcommand{\arraystretch}{1.5} 
\centering
\begin{tabular}{l l l}
\hline
Parameter [units] & First value & Second value  \\ 
\hline
$\kappa$ [g ml$^{-1}$s$^{-1}$] & 5.8 & 5.4 \\
$\eta$ [Nm$^{-2}$] & 5.3 & 1.9 \\
$\nu$ [Nm$^{-2}$] & 5.9 & 5.9 \\
$\delta_n$ [Nm$^{-2}$] & 5.6 & 5.5 \\
\hline
\end{tabular}
\caption{The average number of clusters formed over 10 initial conditions at $t$ = 2, when each additional parameter is added in turn using the parameter values presented in Table \ref{tab:1Dadd_dim_param_bounds}. The number of clusters formed without any additional parameters was 5.9 (Table \ref{tab:1Dadd_dim_results}).}
\label{tab:1Dadd_dim_results}
\end{table}



\vspace{-0.5cm}

\subsection{Impact of VEGF-matrix binding mechanism on patterning}
\label{res:vegf_binding}

Motivated by reducing the spatial scale of pattern formation, the inclusion of VEGF-matrix binding was also considered, by utilising Equations (\ref{eqn:2Dn} - \ref{eqn:2Dvw}, \ref{eqn:2Dun_binding} - \ref{eqn:2Dcb}). Given that parameter values for the binding rates $k_b$ and $k_u$ are fixed, based on \cite{Kohn-Luque2013}, this model  including binding was treated as an alternate version to the core model analysed so far, and may require different core parameter values.

\begin{figure}[H]
    \includegraphics[width=1.1\textwidth]{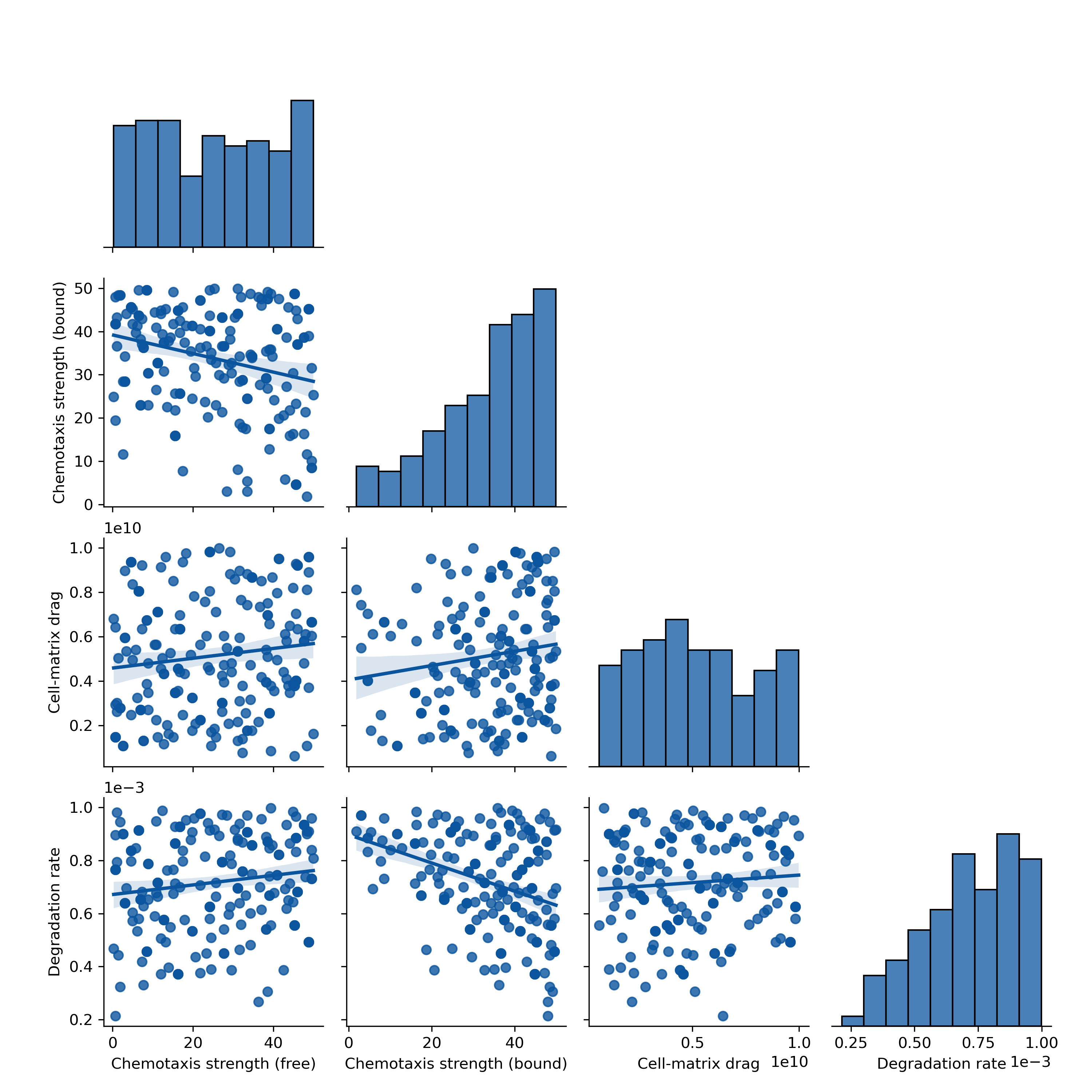}
    \caption{Pairplot illustration of the subset of parameter samples that were found to produce four or more clusters at $t$ = 2 in the VEGF binding model. The histograms present the distribution of the values of each parameter present in the subset; the off-diagonal scatter plots illustrate correlations between parameter values. A regression line is plotted for each off-diagonal scatter plot. Each quantity presented is dimensionless.}
    \label{fig:ps_binding_pairplot}
\end{figure}

A parameter sweep was first conducted: a Saltelli sampler was used to generate 896 parameter samples using the parameter bounds given in Table \ref{tab:binding_param_bounds}. 
The VEGF-matrix binding model achieved a much higher average number of clusters, up to 10, compared to the maximum average of 4.7 observed in results from the original core model.   
Figure \ref{fig:ps_binding_pairplot} presents the subset of the parameter samples in which at least 4 clusters formed at $t = 2$. 

\begin{table}[h]
\setlength{\tabcolsep}{10pt} 
\renewcommand{\arraystretch}{1.5} 
\centering
\begin{tabular}{l l l}
\hline
Parameter [units] & Minimum value & Maximum value  \\ 
\hline
$\chi_u$ [Nm$^{-2}$] & $0$ & $50$ \\
$\chi_b$ [Nm$^{-2}$] & $0$ & $50$ \\
$\gamma_{nm}$ [Nm$^{-4}s$] & $1 \times 10^{8}$ & $1 \times 10^{10}$ \\
$\alpha$ [g ml$^{-1}$s$^{-1}$] & $1 \times 10^{-11}$ & $9 \times 10^{-11}$ \\
$\delta$ [s$^{-1}$]  & $1 \times 10^{-4}$ & $1 \times 10^{-3}$ \\
\hline
\end{tabular}
\caption{Dimensional parameter bounds for the VEGF-matrix binding model parameters used for parameter sweep.}
\label{tab:binding_param_bounds}
\end{table}

The histograms on the diagonal of Figure \ref{fig:ps_binding_pairplot} indicate the spread of parameter values in the data subset. In comparison to the core model in Figure \ref{fig:ps_pairplot}, the range of values of free chemotaxis strength and cell-matrix drag given by the histograms are relatively well-spread, indicating that in the presence of VEGF-matrix binding, there is less of a dependence on free chemotaxis strength and cell-matrix drag in terms of cell cluster formation. The range of values of bound chemotaxis strength in this dataset is strongly skewed towards higher values, suggesting that a strong chemotactic response to bound VEGF is strongly linked to the number of clusters formed. Although the histogram for degradation rate is skewed similarly as for the core model, we can see from the regression plot between chemotaxis strength (bound) and VEGF degradation rate that, as the bound chemotaxis strength increases, there is a broader range of degradation rates present in the dataset. This indicates that unlike in the core model, the degradation rate does not strictly need to be at the upper limit when chemotaxis to bound VEGF is strong. 

To verify the maximum number of cell clusters that can be formed in the 1D geometry with VEGF-matrix binding, a parameter optimisation using particle swarm optimisation was repeated. The optimised parameters found to maximise the number of clusters forming, averaged over 10 initial conditions as in the core model analysis, are given for $t = 2$ and $t = 4$ in Table \ref{tab:po_1D_binding}. At both $t = 2$ and $t = 4$, the optimum parameter values were found to be for relatively low chemotaxis strength to unbound VEGF, and relatively high chemotaxis strength to bound VEGF. Moreover, the parameter values found for VEGF production and degradation rate for $t = 4$ indicate that high values for these parameters are not strictly necessary for a large number of clusters to form, which further supports the findings of the parameter sweep in Figure \ref{fig:ps_binding_pairplot}.

\begin{table}[h]
\setlength{\tabcolsep}{10pt} 
\renewcommand{\arraystretch}{1.5} 
\centering
\begin{tabular}{l l l l l}
\hline
$t$ [-] & 2 & 4 \\
\hline
average no. of clusters & $9.3$ & $8.0$ \\ 
VEGF degradation rate & $8.21 \times 10^{-4}$ & $3.32 \times 10^{-4}$ \\
VEGF production rate & $8.74 \times 10^{-11}$ & $4.11 \times 10^{-11}$ \\
Cell-matrix drag & $9.89 \times 10^9$ & $7.57 \times 10^9$ \\
Chemotaxis strength (bound) & $42.4$ & $49.7$ \\
Chemotaxis strength (unbound) & $1.36$ & $9.04$  \\
\hline
\hline
\end{tabular}
\caption{Parameter values found to maximise the number of cell clusters produced by the VEGF-matrix binding model at $t = 2$ and $t = 4$ respectively, using the particle swarm optimisation method.}
\label{tab:po_1D_binding}
\end{table}

The findings of the above parameter sweep and particle swarm optimisation appear to suggest that including VEGF-matrix binding leads to a more robust simulation outcome against variation in VEGF production and degradation rates. To investigate this hypothesis, a sensitivity analysis was performed on the VEGF-matrix binding model parameters based on the chemotaxis, cell-matrix drag, production and degradation upper and lower bounds given in Table \ref{tab:binding_param_bounds}, as well as the upper and lower bounds for the rate of binding and unbinding given by K\"ohn-Luque et al.~\cite{Kohn-Luque2013}. The results of the sensitivity analysis are shown in Table \ref{tab:sa_binding}, where the influence of the strength of chemotaxis to bound VEGF at $t=4$ is higher than for the strength of chemotaxis to free VEGF. All other parameters exhibit a similar influence as in the core model, though the Si values for VEGF production and degradation are lower than those obtained for the core model (Table \ref{tab:sa_coreparameters}) as expected. The low Si values for the rates of binding and unbinding confirm that the experimental upper and lower bounds are sufficiently narrow that the uncertainty in these parameters do not overshadow the impact of the other model parameters on the model outcome. 

\begin{table}[H]
\setlength{\tabcolsep}{10pt} 
\renewcommand{\arraystretch}{1.5} 
\centering
\begin{tabular}{l l l l l}
\hline
$t$ [-] & $1$ & $2$ & $3$ & $4$  \\
\hline
VEGF degradation rate & $0.63$ & $0.61$ & $0.63$ & $0.64$ \\
VEGF production rate & $0.49$ & $0.46$ & $0.45$ & $0.41$ \\
Chemotaxis strength (free VEGF) & $0.52$ & $0.32$ & $0.26$ & $0.26$ \\
Chemotaxis strength (bound VEGF) & $0.50$ & $0.39$ & $0.36$ & $0.32$ \\
Cell-matrix drag & $0.23$ & $0.24$ & $0.22$ & $0.19$ \\
Rate of binding & $0.08$ & $0.08$ & $0.09$ & $0.09$ \\
Rate of unbinding & $0.11$ & $0.12$ & $0.12$ & $0.13$ \\
\hline
\hline
\end{tabular}
\caption{Total Sensitivity indices for each of the parameters in the extended VEGF-matrix binding model, based on the number of cell clusters formed, at $t = 1$, $2$, $3$, and $4$ respectively.}
\label{tab:sa_binding}
\end{table}

\begin{figure}[H]
\hfill
\begin{subfigure}{0.46\textwidth}
    \centering
    \includegraphics[width = \textwidth]{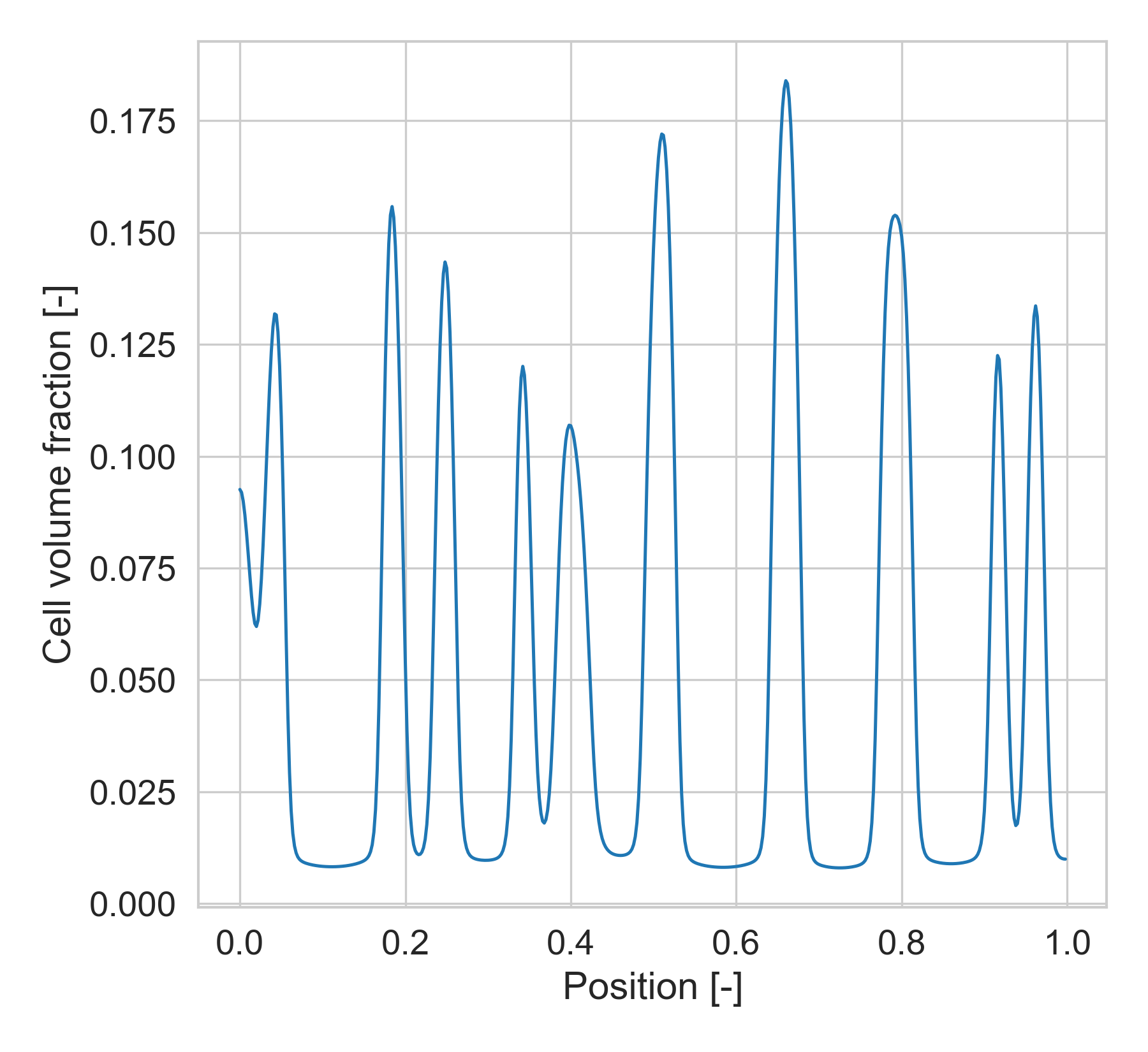}
    \caption{Example cell cluster formation in 1D}
    \label{fig:po_binding_1D_example}
\end{subfigure}
\hfill
\begin{subfigure}{0.52\textwidth}
    \centering
    \includegraphics[width = \textwidth]{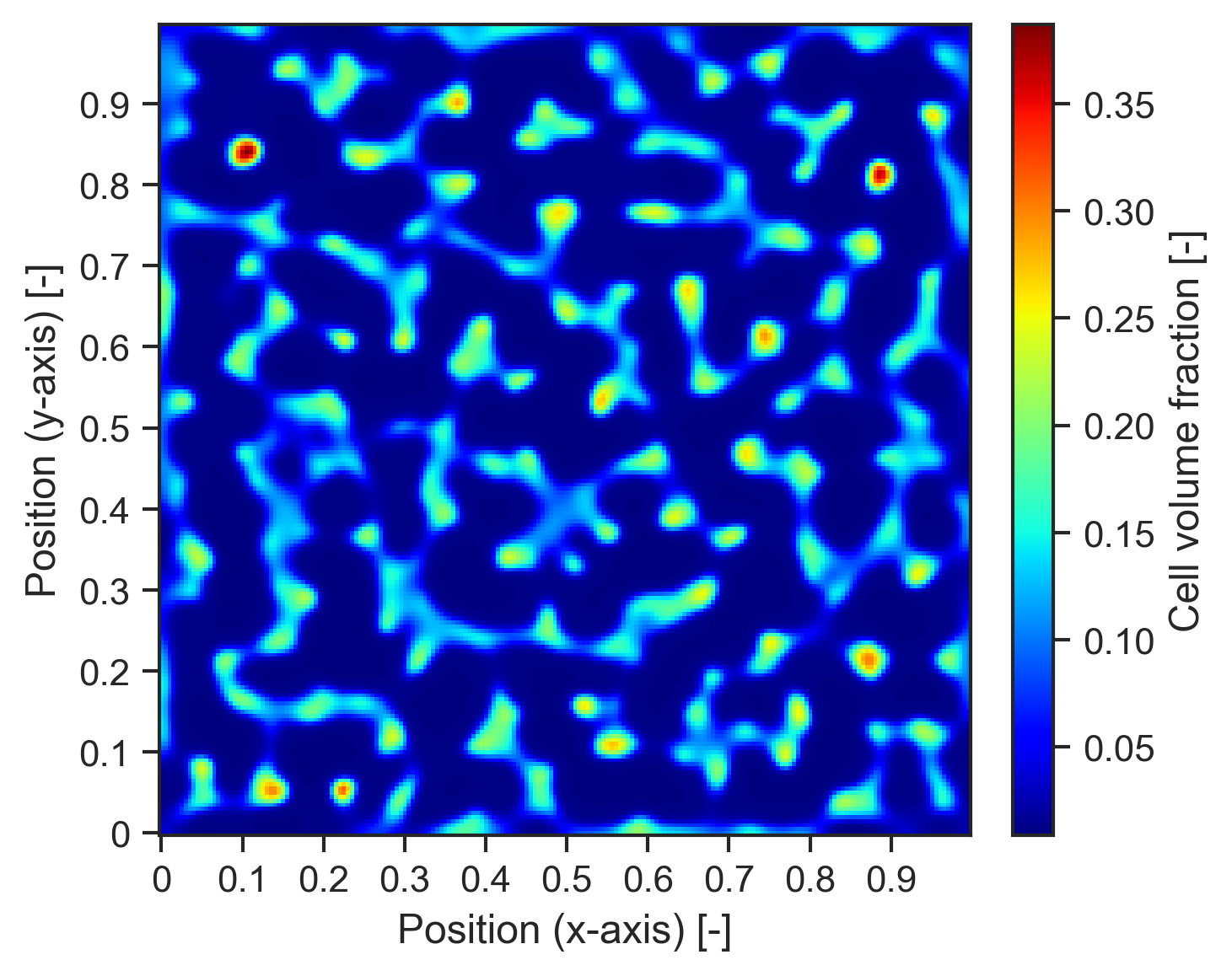}
    \caption{Example cell pattern formation in 2D}
    \label{fig:po_binding_2D_example}
\end{subfigure}
\caption{Example cell cluster and pattern formation for the VEGF-matrix binding model with the optimised parameter values given in the first column ($t$ = 2) of Table \ref{tab:po_1D_binding}, shown at $t$ = 2. The same random initial condition was used to generate all 1D and 2D example plots.}
\label{fig:po_binding_examples}
\end{figure}

Figure \ref{fig:po_binding_examples} illustrates cell clustering produced by the VEGF-matrix binding model using the optimised parameter values in Table \ref{tab:po_1D_binding}. Figure \ref{fig:po_binding_2D_example} both confirms the ability of the extended VEGF-matrix binding model to generate connected vascular-like structures on a biologically relevant timescale, and indicates that the VEGF-matrix binding model can better match the spatial scale of vascular network formation as indicated by the 1D analysis. 

\subsection{Comparison of spatial scale of patterning}
\label{res:spac_anal_all}

\begin{figure}[H]
\hfill
\begin{subfigure}{0.49\textwidth}
    \centering
    \includegraphics[width = \textwidth]{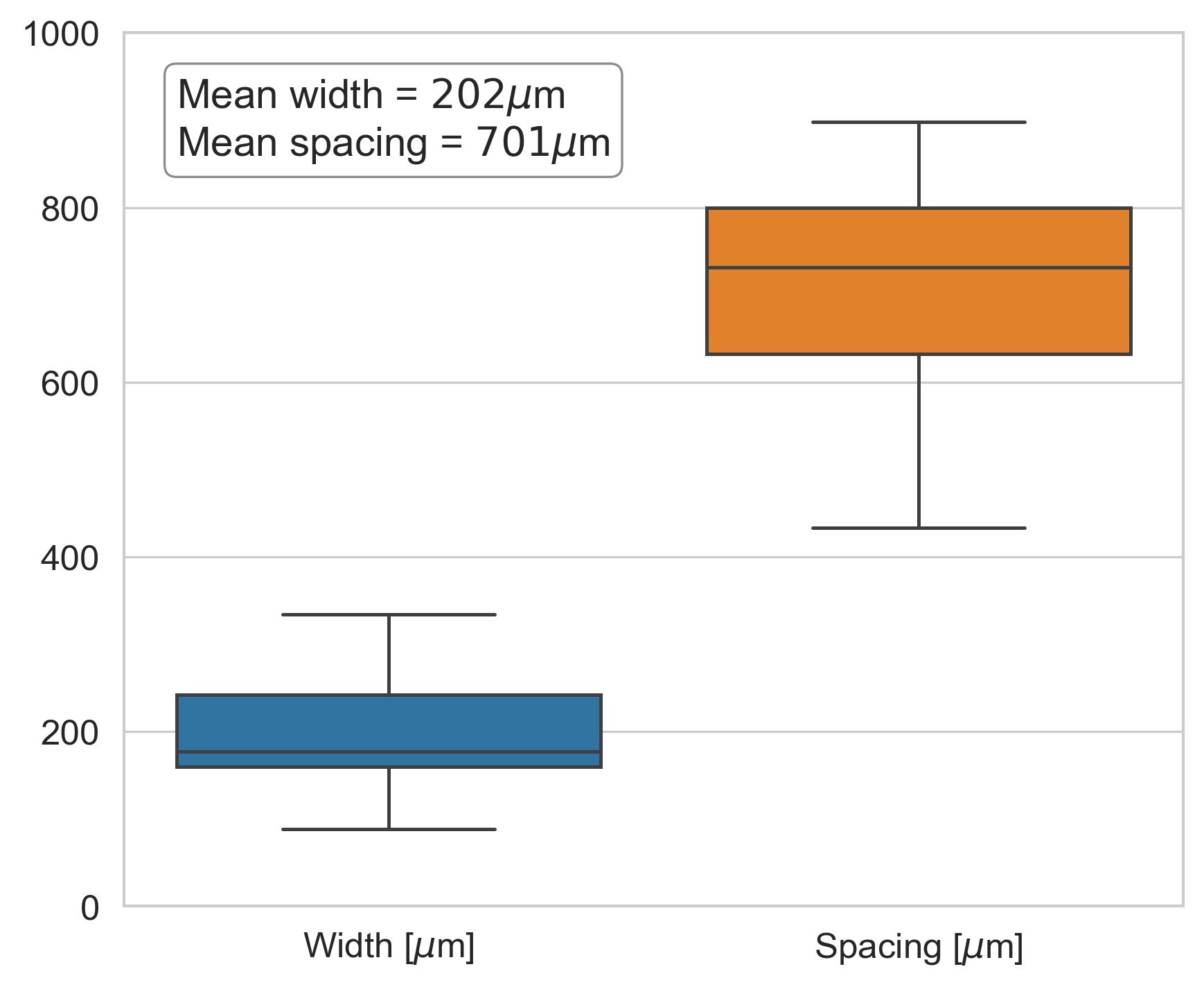}
    \caption{1D core model}
    \label{fig:metrics_1D_core}
\end{subfigure}
\hfill
\begin{subfigure}{0.49\textwidth}
    \centering
    \includegraphics[width = \textwidth]{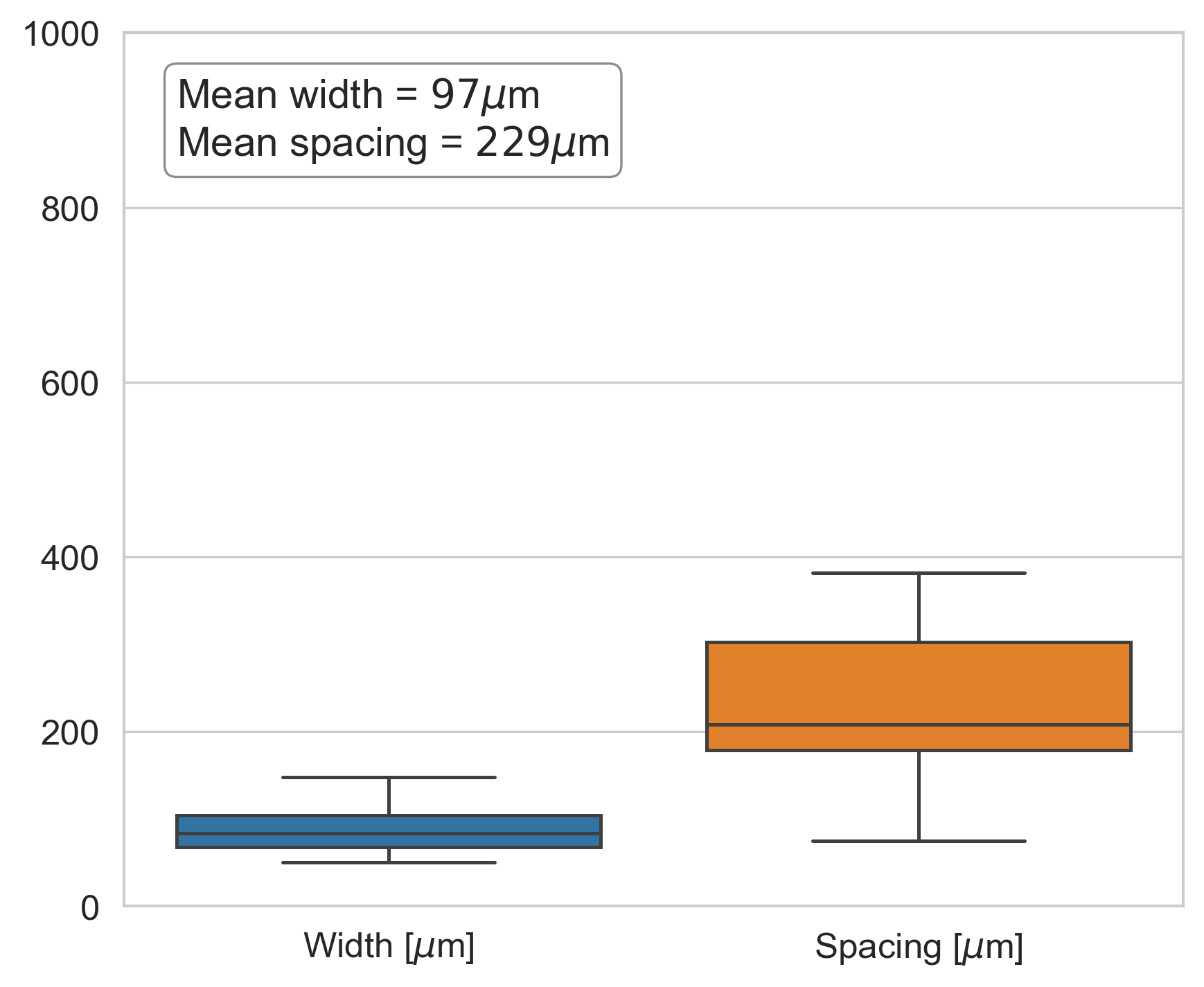}
    \caption{1D binding model}
    \label{fig:metrics_1D_binding}
\end{subfigure}
\caption{Spatial metrics for 1D core and 1D binding model, using optimised parameter values. The metrics were extracted from 5 simulations given random initial conditions.}
\label{fig:metrics_1D}
\end{figure}

To confirm the typical spatial scale of the core and binding models, assessed by width and spacing of each cell structure formed, these metrics were assessed for each model over 5 random initial conditions. In Figure \ref{fig:metrics_1D}, the distribution of metrics in the 1D core model is compared with the metrics in the 1D binding model. This demonstrates a clear reduction in both metrics, with a significant different in cluster spacing. For the 1D binding model, the average cluster spacing was found to be 229 $\mu$m, of the same order of vessel separation observed \textit{in vitro} \cite{Serini2003}. 

\begin{figure}[H]
\hfill
\begin{subfigure}{0.49\textwidth}
    \centering
    \includegraphics[width = \textwidth]{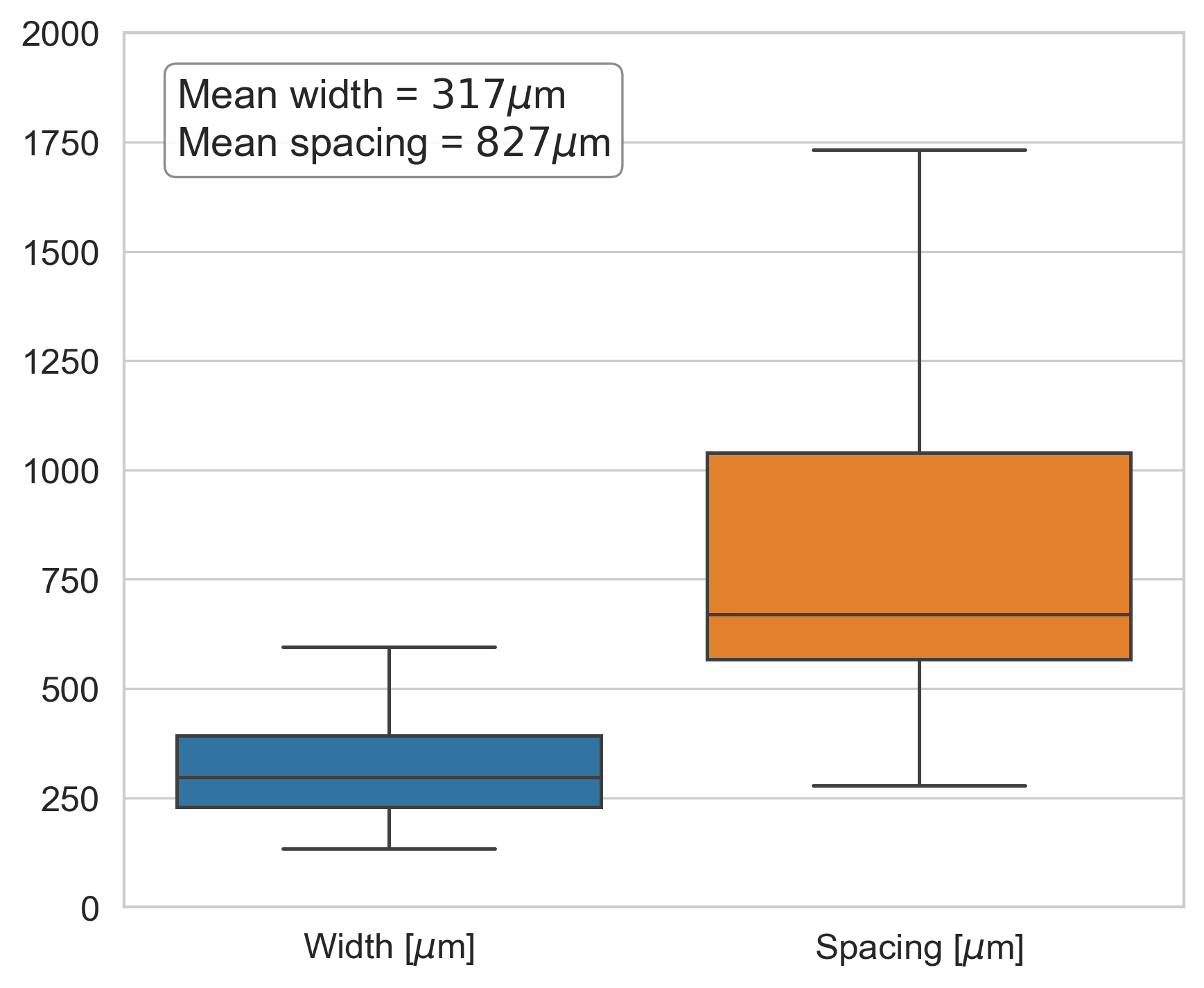}
    \caption{2D core model}
    \label{fig:metrics_1D_core}
\end{subfigure}
\hfill
\begin{subfigure}{0.49\textwidth}
    \centering
    \includegraphics[width = \textwidth]{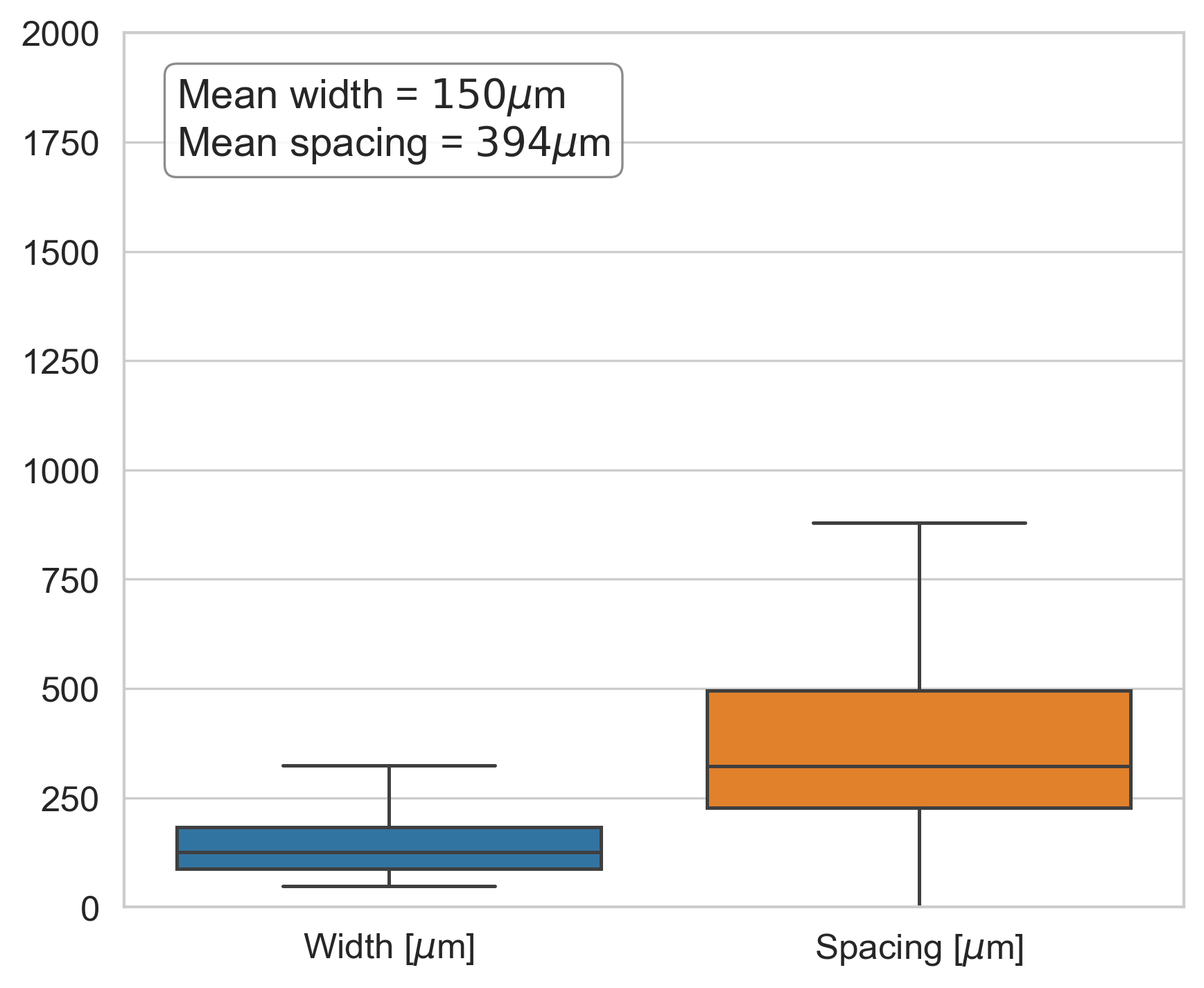}
    \caption{2D binding model}
    \label{fig:metrics_2D_binding}
\end{subfigure}
\caption{Spatial metrics for 2D core and 2D binding model, using optimised parameter values. The metrics were extracted from 5 simulations given random initial conditions.}
\label{fig:metrics_2D}
\end{figure}

Figure \ref{fig:metrics_2D} makes a similar comparison for the 2D core and 2D binding models, with clear reduction in both metrics for the binding model. Compared to 1D, each metric average appears slightly higher, likely due to the method employed to determine these metrics. In 2D, the metrics were determined by taking several 1D slices in each dimension, however this does not guarantee to pass through the most narrow part of each cell structure, nor capture the smallest separation of cell structures, illustrated by the much larger spread of measurements for the 2D metrics. A more sophisticated method to capture the spacing as related to the minor axis of each void of the 2D vessel-like structure would be required to assert that the spacing of the 2D model is also of order of 200 $\mu$m.

\section{Conclusion}
\label{sec:conclusion}

In this paper, we have demonstrated the potential for the multiphase model to be used to simulate cell patterning in 1D and 2D, including both cell cluster formation and the formation of vascular-like structures. We have investigated cell mechanisms commonly included in models of vascular network formation models, and more general multiphase models, with a focus on the influence of each parameter on the spatial and temporal scale of patterning. Using sensitivity analysis, 4 core parameters were identified and demonstrated to enable the formation of cell clusters and vascular-like structures in 1D and 2D respectively, on a biologically plausible timescale (2 to 4 days). Additional model parameters, including cell-matrix traction, VEGF uptake, cell-cell aggregation, and cell contact inhibition, were not found to favourably augment the spatial scale of pattern formation. Alternatively, the addition of VEGF-matrix binding as a model extension was found to match the spatial scale of patterning as observed \textit{in vitro}, with significantly smaller pattern spacing compared to the core model.

In this analysis, we have demonstrated that the combination of analyses methods employed here: parameter sweep, sensitivity analysis, and parameter optimisation, is an efficient way to guide the optimisation of the 2D vascular network formation model. Further, the sensitivity analysis identified an increase in robustness of the binding model to some of the core parameters, particularly for VEGF production and VEGF degradation.  

Future analysis may include the application of some of these computational techniques, including parameter optimisation, to the 2D model directly. Additionally, the analysis would benefit from a deeper look at persistence of the vascular-like structures in 2D, and the stability of metrics observed over a longer time point. To adapt the model presented here to long-term behaviour, it is likely that additional stabilising mechanisms such as contact inhibition, or a temporal-dependence on the cell-matrix interactions, such as by cell-driven augmentation of the matrix distribution, may be required to match \textit{in vitro} findings. 

Overall, this study has demonstrated potential for the multiphase model to capture key \textit{in vitro} behaviours at the time points considered, which makes it a prime candidate for further development to be utilised alongside experimental studies. In particular, the wide range of experimental variables associated with \textit{in vitro} vascular network formation can be assesses computationally by sensitivity analysis, to determine the most influential. The combination with computational analysis, such as demonstrated here, has the power to significantly reduce the experimental load towards \textit{in vitro} development of vascularised engineered tissues and accelerate research in this area towards clinical application. 

\section*{Acknowledgements}

We thank Maxime Berg for useful discussions surrounding the computational implementation of the model. This work was supported by EPSRC grants EP/R512400/1 and EP/W522636/1.



\bibliographystyle{unsrt}
\bibliography{elsarticle-template-num}

\begin{thebibliography}{10}

\bibitem{Bittner2020}
Katharine~R. Bittner, Juan~M. Jim{\'{e}}nez, and Shelly~R. Peyton.
\newblock {Vascularized Biomaterials to Study Cancer Metastasis}, apr 2020.

\bibitem{Tremblay2005}
Pierre~Luc Tremblay, Val{\'{e}}rie Hudon, Fran{\c{c}}ois Berthod, Lucie
  Germain, and Fran{\c{c}}ois~A. Auger.
\newblock {Inosculation of tissue-engineered capillaries with the host's
  vasculature in a reconstructed skin transplanted on mice}.
\newblock {\em American Journal of Transplantation}, 5(5):1002--1010, 2005.

\bibitem{Shen2016}
Yu~I. Shen, Hongkwan Cho, Arianne~E. Papa, Jacqueline~A. Burke, Xin~Yi Chan,
  Elia~J. Duh, and Sharon Gerecht.
\newblock {Engineered human vascularized constructs accelerate diabetic wound
  healing}.
\newblock {\em Biomaterials}, 102:107--119, sep 2016.

\bibitem{Song2018}
H.~H.Greco Song, Rowza~T. Rumma, C.~Keith Ozaki, Elazer~R. Edelman, and
  Christopher~S. Chen.
\newblock {Vascular Tissue Engineering: Progress, Challenges, and Clinical
  Promise}, mar 2018.

\bibitem{Yang2020}
Guang Yang, Bhushan Mahadik, Ji~Young Choi, and John~P Fisher.
\newblock {Vascularization in tissue engineering: fundamentals and
  state-of-art}.
\newblock {\em Progress in Biomedical Engineering}, 2(1):012002, jan 2020.

\bibitem{Rademakers2019}
Timo Rademakers, Judith~M. Horvath, Clemens~A. van Blitterswijk, and
  Vanessa~L.S. LaPointe.
\newblock {Oxygen and nutrient delivery in tissue engineering: Approaches to
  graft vascularization}, oct 2019.

\bibitem{Chang2017}
William~G. Chang and Laura~E. Niklason.
\newblock {A short discourse on vascular tissue engineering}.
\newblock {\em npj Regenerative Medicine}, 2(1), dec 2017.

\bibitem{Scianna2013}
M~Scianna, C~G Bell, and L~Preziosi.
\newblock {A review of mathematical models for the formation of vascular
  networks}.
\newblock {\em Journal of theoretical biology}, 333:174--209, 2013.

\bibitem{Waters2021}
S.~L. Waters, L.~J. Schumacher, and A.~J. {El Haj}.
\newblock {Regenerative medicine meets mathematical modelling: developing
  symbiotic relationships}, apr 2021.

\bibitem{ODea2012}
RD~O'Dea, HM~Byrne, and SL~Waters.
\newblock {Continuum Modelling of In Vitro Tissue Engineering: A Review}.
\newblock In {\em Studies in Mechanobiology, Tissue Engineering and
  Biomaterials}, volume~10, pages 229--266. Springer, 2012.

\bibitem{Lemon2006}
Greg Lemon, John~R. King, Helen~M. Byrne, Oliver~E. Jensen, and Kevin~M.
  Shakesheff.
\newblock {Mathematical modelling of engineered tissue growth using a
  multiphase porous flow mixture theory}.
\newblock {\em Journal of Mathematical Biology}, 52(5):571--594, 2006.

\bibitem{Lemon2007}
Greg Lemon and John~R King.
\newblock {Multiphase modelling of cell behaviour on artificial scaffolds:
  Effects of nutrient depletion and spatially nonuniform porosity}.
\newblock {\em Mathematical Medicine and Biology}, 24(1):57--83, 2007.

\bibitem{Byrne2003}
Helen Byrne and Luigi Preziosi.
\newblock {Modelling solid tumour growth using the theory of mixtures}.
\newblock {\em Mathematical Medicine and Biology}, 20(4):341--366, 2003.

\bibitem{Preziosi2009}
Luigi Preziosi and Andrea Tosin.
\newblock {Multiphase modelling of tumour growth and extracellular matrix
  interaction: Mathematical tools and applications}.
\newblock {\em Journal of Mathematical Biology}, 58(4-5):625--656, 2009.

\bibitem{Tosin2010}
Andrea Tosin and Luigi Preziosi.
\newblock {Multiphase modeling of tumor growth with matrix remodeling and
  fibrosis}.
\newblock {\em Mathematical and Computer Modelling}, 52(7-8):969--976, oct
  2010.

\bibitem{Hubbard2013}
M.~E. Hubbard and H.~M. Byrne.
\newblock {Multiphase modelling of vascular tumour growth in two spatial
  dimensions}.
\newblock {\em Journal of Theoretical Biology}, 316:70--89, jan 2013.

\bibitem{Sciume2013}
G~Scium{\`{e}}, S~Shelton, W~G Gray, C~T Miller, F~Hussain, M~Ferrari,
  P~Decuzzi, and B~A Schrefler.
\newblock {A multiphase model for three-dimensional tumor growth}.
\newblock {\em New Journal of Physics}, 15(35pp):15005, 2013.

\bibitem{Dyson2016}
R~J Dyson, J~E {F Green}, J~P Whiteley, H~M Byrne, R~J {Dyson RJDyson},
  bhamacuk~J {P Whiteley JonathanWhiteley}, and csoxacuk~H {M Byrne
  HelenByrne}.
\newblock {An investigation of the influence of extracellular matrix anisotropy
  and cell–matrix interactions on tissue architecture}.
\newblock {\em J. Math. Biol}, 72:1775--1809, 2016.

\bibitem{Green2017}
J.~E.F. Green, J~P Whiteley, J~M Oliver, H~M Byrne, and S~L Waters.
\newblock {Pattern formation in multiphase models of chemotactic cell
  aggregation}.
\newblock {\em Mathematical medicine and biology : a journal of the IMA},
  35(3):319--346, 2018.

\bibitem{Bayless2000}
K~J Bayless, Ren{\'{e}} Salazar, and George~E Davis.
\newblock {RGD-dependent vacuolation and lumen formation observed during
  endothelial cell morphogenesis in three-dimensional fibrin matrices involves
  the alpha(v)beta(3) and alpha(5)beta(1) integrins.}
\newblock {\em The American journal of pathology}, 156(5):1673--83, 2000.

\bibitem{Koh2008}
Wonshill Koh, Amber~N Stratman, Anastasia Sacharidou, and George~E Davis.
\newblock {In Vitro Three Dimensional Collagen Matrix Models of Endothelial
  Lumen Formation During Vasculogenesis and Angiogenesis}, 2008.

\bibitem{Stratman2009}
Amber~N Stratman, W~Brian Saunders, Anastasia Sacharidou, Wonshill Koh, Kevin~E
  Fisher, David~C Zawieja, Michael~J Davis, and George~E Davis.
\newblock {Endothelial cell lumen and vascular guidance tunnel formation
  requires MT1-MMP–dependent proteolysis in 3-dimensional collagen matrices}.
\newblock {\em Blood}, 114:237--247, 2009.

\bibitem{Blinder2016}
Yaron~J. Blinder, Alina Freiman, Noa Raindel, David~J. Mooney, and Shulamit
  Levenberg.
\newblock {Vasculogenic dynamics in 3D engineered tissue constructs}.
\newblock {\em Scientific Reports}, 5(1):17840, 2016.

\bibitem{Carmeliet1996}
Peter Carmeliet, Val{\'{e}}rie Ferreira, Georg Breier, Saskla Pollefeyt, Lena
  Kieckens, Marina Gertsenstein, Michaela Fahrig, Ann Vandenhoeck, Kendraprasad
  Harpal, Carmen Eberhardt, Cath{\'{e}}rine Declercq, Judy Pawling, Lieve
  Moons, D{\'{e}}sir{\'{e}} Collen, Werner Risaut, and Andras Nagy.
\newblock {Abnormal blood vessel development and lethality in embryos lacking a
  single VEGF allele}.
\newblock {\em Nature}, 380(6573):435--439, 1996.

\bibitem{Ferrara1996}
Napoleone Ferrara, Karen Carver-Moore, Helen Chen, Mary Dowd, Lucy Lu, K.~Sue
  O'Shea, Lyn Powell-Braxton, Kenneth~J Hillan, and Mark~W Moore.
\newblock {Heterozygous embryonic lethality induced by targeted inactivation of
  the VEGF gene}.
\newblock {\em Nature}, 380:439--442, 1996.

\bibitem{Zetter1980}
Bruce~R. Zetter.
\newblock {Migration of capillary endothelial cells is stimulated by
  tumour-derived factors.}
\newblock {\em Nature}, 285(5759):41--3, 1980.

\bibitem{Barkefors2008}
Irmeli Barkefors, S{\'{e}}bastien {Le Jan}, Lars Jakobsson, Eduar Hejll, Gustav
  Carlson, Henrik Johansson, Jonas Jarvius, Won~Park Jeong, Li~Jeon Noo, and
  Johan Kreuger.
\newblock {Endothelial cell migration in stable gradients of vascular
  endothelial growth factor A and fibroblast growth factor 2: Effects on
  chemotaxis and chemokinesis}.
\newblock {\em Journal of Biological Chemistry}, 283(20):13905--13912, 2008.

\bibitem{Wu2014}
Pian Wu, Ya~Fu, and Kaiyong Cai.
\newblock {Regulation of the migration of endothelial cells by a gradient
  density of vascular endothelial growth factor}.
\newblock {\em Colloids and Surfaces B: Biointerfaces}, 123:181--190, 2014.

\bibitem{Unemori1992}
Elaine~N. Unemori, Napoleone Ferrara, Eugene~A. Bauer, and Edward~P. Amento.
\newblock {Vascular endothelial growth factor induces interstitial collagenase
  expression in human endothelial cells}.
\newblock {\em Journal of Cellular Physiology}, 153(3):557--562, dec 1992.

\bibitem{Seghezzi1998}
Graziano Seghezzi, Sundeep Patel, Christine~J. Ren, Anna Gualandris, Giuseppe
  Pintucci, Edith~S. Robbins, Richard~L. Shapiro, Aubrey~C. Galloway, Daniel~B.
  Rifkin, and Paolo Mignatti.
\newblock {Fibroblast growth factor-2 (FGF-2) induces vascular endothelial
  growth factor (VEGF) expression in the endothelial cells of forming
  capillaries: An autocrine mechanism contributing to angiogenesis}.
\newblock {\em Journal of Cell Biology}, 141(7):1659--1673, jun 1998.

\bibitem{Hanjaya-Putra2010}
Donny Hanjaya-Putra, Jane Yee, Doug Ceci, Rachel Truitt, Derek Yee, and Sharon
  Gerecht.
\newblock {Vascular endothelial growth factor and substrate mechanics regulate
  in vitro tubulogenesis of endothelial progenitor cells}.
\newblock {\em Journal of Cellular and Molecular Medicine}, 14(10):2436--2447,
  oct 2010.

\bibitem{Helmlinger2000}
Gabriel Helmlinger, Mitsuhiro Endo, Napoleone Ferrara, Lynn Hlatky, and
  Rakesh~K. Jain.
\newblock {Growth factors: Formation of endothelial cell networks}.
\newblock {\em Nature}, 405(6783):139--141, may 2000.

\bibitem{Serini2003}
G~Serini, D~Ambrosi, E~Giraudo, A~Gamba, L~Preziosi, and F~Bussolino.
\newblock {Modeling the early stages of vascular network assembly}.
\newblock {\em Embo J}, 22(8):1771--1779, 2003.

\bibitem{Macri2007}
Lauren Macri, David Silverstein, and Richard~A.F. Clark.
\newblock {Growth factor binding to the pericellular matrix and its importance
  in tissue engineering}.
\newblock {\em Advanced Drug Delivery Reviews}, 59(13):1366--1381, 2007.

\bibitem{Wijelath2006}
Errol~S Wijelath, Salman Rahman, Mayumi Namekata, Jacqueline Murray, and
  Tomoaki Nishimura.
\newblock {Heparin-II Domain of Fibronectin Is a Vascular Endothelial Growth
  Factor–Binding Domain}.
\newblock {\em Circulation Research}, 99:853--860, dec 2006.

\bibitem{Sahni2000}
Abha Sahni, Charles~W Francis, and Washington Dc.
\newblock {Vascular endothelial growth factor binds to fibrinogen and fibrin
  and stimulates endothelial cell proliferation Vascular endothelial growth
  factor binds to fibrinogen and fibrin and stimulates endothelial cell
  proliferation}.
\newblock {\em Blood}, 96(12):3772--3778, 2000.

\bibitem{Ruhrberg2002}
Christiana Ruhrberg, Holger Gerhardt, Matthew Golding, Rose Watson, Sofia
  Ioannidou, Hajime Fujisawa, Christer Betsholtz, and David~T. Shima.
\newblock {Spatially restricted patterning cues provided by heparin-binding
  VEGF-A control blood vessel branching morphogenesis}.
\newblock {\em Genes and Development}, 16(20):2684--2698, oct 2002.

\bibitem{Kohn-Luque2013}
Alvaro K{\"{o}}hn-Luque, W~de~Back, Y~Yamaguchi, K~Yoshimura, M~A Herrero, and
  T~Miura.
\newblock {Dynamics of VEGF matrix-retention in vascular network patterning}.
\newblock {\em Physical Biology}, 10(6):066007, 2013.

\bibitem{Tosin2006}
A~Tosin, D~Ambrosi, and L~Preziosi.
\newblock {Mechanics and chemotaxis in the morphogenesis of vascular networks}.
\newblock {\em Bulletin of Mathematical Biology}, 68(7):1819--1836, 2006.

\bibitem{Manoussaki1996}
D~Manoussaki, S~R Lubkin, R~B Vernon, and J~D Murray.
\newblock {A mechanical model for the formation of vascular networks in vitro}.
\newblock {\em Acta biotheoretica}, 44(3-4):271--282, 1996.

\bibitem{Tranqui2000}
L{\'{e}}one Tranqui and Philippe Tracqui.
\newblock {Mechanical signalling and angiogenesis. The integration of
  cell–extracellular matrix couplings}.
\newblock {\em Comptes Rendus de l'Acad{\'{e}}mie des Sciences - Series III -
  Sciences de la Vie}, 323(1):31--47, 2000.

\bibitem{Namy2004}
Patrick Namy, Jacques Ohayon, and Philippe Tracqui.
\newblock {Critical conditions for pattern formation and in vitro tubulogenesis
  driven by cellular traction fields}.
\newblock {\em Journal of Theoretical Biology}, 227(1):103--120, 2004.

\bibitem{ODea2009}
R.~D. O'Dea, S.~L. Waters, and H.~M. Byrne.
\newblock {A multiphase model for tissue construct growth in a perfusion
  bioreactor}.
\newblock {\em Mathematical Medicine and Biology}, 27(2):95--127, 2009.

\bibitem{Pearson2014}
Natalie~C. Pearson, Rebecca~J. Shipley, Sarah~L. Waters, and James~M. Oliver.
\newblock {Multiphase modelling of the influence of fluid flow and chemical
  concentration on tissue growth in a hollow fibre membrane bioreactor}.
\newblock {\em Mathematical Medicine and Biology}, 31(4):393--430, 2014.

\bibitem{Odedra2011}
Devang Odedra, Loraine L~Y Chiu, Molly Shoichet, and Milica Radisic.
\newblock {Endothelial cells guided by immobilized gradients of vascular
  endothelial growth factor on porous collagen scaffolds}.
\newblock {\em Acta Biomaterialia}, 7(8):3027--3035, 2011.

\bibitem{Holmes2000}
M.J. Holmes and B.D. Sleeman.
\newblock {A Mathematical Model of Tumour Angiogenesis Incorporating Cellular
  Traction and Viscoelastic Effects}.
\newblock {\em Journal of Theoretical Biology}, 202(2):95--112, 2000.

\bibitem{Paponthesis}
Papon Muangsanit.
\newblock {\em {Aligned endothelial cell and Schwann cell structures in 3D
  hydrogels for peripheral nerve tissue engineering}}.
\newblock PhD thesis, University College London, 2019.

\bibitem{Stokes1991}
C.~L. Stokes, D.~A. Lauffenburger, and S.~K. Williams.
\newblock {Migration of individual microvessel endothelial cells: Stochastic
  model and parameter measurement}.
\newblock {\em Journal of Cell Science}, 99(2):419--430, 1991.

\bibitem{Rupnick1988}
M~A Rupnick, C~L Stokes, S~K Williams, and D~A Lauffenburger.
\newblock {Quantitative analysis of random motility of human microvessel
  endothelial cells using a linear under-agarose assay.}
\newblock {\em Laboratory investigation; a journal of technical methods and
  pathology}, 59(3):363--72, sep 1988.

\bibitem{Anderson1998c}
A.~R.A. Anderson and M.~A.J. Chaplain.
\newblock {Continuous and discrete mathematical models of tumor-induced
  angiogenesis}.
\newblock {\em Bulletin of Mathematical Biology}, 60(5):857--899, 1998.

\bibitem{Nunez2006}
Daniel~A Nunez.
\newblock {\em {Experimental estimate of the diffusivity of Vascular
  Endothelial Growth Factor}}.
\newblock PhD thesis, Massachusetts Institute of Technology, 2006.

\bibitem{MacGabhann2007}
Feilim {Mac Gabhann}, James~W. Ji, and Aleksander~S. Popel.
\newblock {VEGF gradients, receptor activation, and sprout guidance in resting
  and exercising skeletal muscle}.
\newblock {\em Journal of Applied Physiology}, 102(2):722--734, 2007.

\bibitem{Kohn-Luque2011}
Alvaro K{\"{o}}hn-Luque, Walter de~Back, J{\"{o}}rn Starru{\ss}, Andrea
  Mattiotti, Andreas Deutsch, Jos{\'{e}}~Mar{\'{i}}a P{\'{e}}rez-Pomares, and
  Miguel~A. Herrero.
\newblock {Early embryonic vascular patterning by matrix-mediated paracrine
  signalling: A mathematical model study}.
\newblock {\em PLoS ONE}, 6(9), 2011.

\bibitem{Merks2006}
Roeland M~H Merks and James~A Glazier.
\newblock {Dynamic mechanisms of blood vessel growth.}
\newblock {\em Nonlinearity}, 19(1):C1--C10, 2006.

\bibitem{Chen2010}
Tom~T. Chen, Alfonso Luque, Sunyoung Lee, Sean~M. Anderson, Tatiana Segura, and
  M.~Luisa Iruela-Arispe.
\newblock {Anchorage of VEGF to the extracellular matrix conveys differential
  signaling responses to endothelial cells}.
\newblock {\em Journal of Cell Biology}, 188(4):595--609, 2010.

\bibitem{Nomura1995}
M.~Nomura, S.~I. Yamagishi, S.~I. Harada, Y.~Hayashi, T.~Yamashima,
  J.~Yamashita, and H.~Yamamoto.
\newblock {Possible participation of autocrine and paracrine vascular
  endothelial growth factors in hypoxia-induced proliferation of endothelial
  cells and pericytes}.
\newblock {\em Journal of Biological Chemistry}, 270(47):28316--28324, 1995.

\bibitem{Coy2020}
R.~Coy, G.~Al-Badri, C.~Kayal, C.~O'Rourke, P.~J. Kingham, J.~B. Phillips, and
  R.~J. Shipley.
\newblock {Combining in silico and in vitro models to inform cell seeding
  strategies in tissue engineering}.
\newblock {\em Journal of the Royal Society Interface}, 17(164):20190801, mar
  2020.

\bibitem{Herman2017}
Jon Herman and Will Usher.
\newblock {SALib: An open-source Python library for Sensitivity Analysis}.
\newblock {\em The Journal of Open Source Software}, 2(9):97, jan 2017.

\bibitem{Saltelli2010}
Andrea Saltelli, Paola Annoni, Ivano Azzini, Francesca Campolongo, Marco Ratto,
  and Stefano Tarantola.
\newblock {Variance based sensitivity analysis of model output. Design and
  estimator for the total sensitivity index}.
\newblock {\em Computer Physics Communications}, 181(2):259--270, feb 2010.

\bibitem{Miranda2018}
Lester {James V. Miranda}.
\newblock {PySwarms: a research toolkit for Particle Swarm Optimization in
  Python}.
\newblock {\em The Journal of Open Source Software}, 3(21):433, 2018.

\end{thebibliography}





\end{document}